\documentclass[twoside,11pt]{article}

\usepackage[accepted,specialissue]{melba}
\usepackage{amsmath,amsfonts}
\usepackage{multirow}
\usepackage{tikz}
\usepackage{pifont}  
\usepackage{caption}
\usepackage{subcaption}
\usepackage{soul}
\usepackage{diagbox}
\usepackage{xcolor}
\usepackage{mathtools}
\usepackage{comment}
\usepackage{colortbl}
\usepackage{bm}

\newcommand*\circled{\tikz[baseline=(current bounding box.center)]{
            \node[shape=circle,draw,inner sep=1.2pt] (char) {};}}

\makeatletter
\newcommand{\labeltext}[2]{%
  \@bsphack
  \def\@currentlabel{#2}{\label{#1}}%
  \csname phantomsection\endcsname 
  \@esphack
}
\makeatother

\usepackage{nameref,zref-xr}
\zxrsetup{toltxlabel}

\melbaid{2024:004}  
\doi{https://doi.org/10.59275/j.melba.2024-d61g}
\melbaauthors{Chen, Staring, Neve, Romeijn, Hensen, Verbist, Wolterink, Tao}  
\volume{2}
\firstpageno{657}  
\melbayear{2024}  
\datesubmitted{08/2023}  
\datepublished{02/2024}  

\melbaspecialissue{Generative Models}
\melbaspecialissueeditors{Mert Sabuncu, Sotirios A. Tsaftaris}

\ShortHeadings{Conditional neural fields with shift modulation}{Chen, Staring, Neve, Romeijn, Hensen, Verbist, Wolterink and Tao}

\title{CoNeS: Conditional neural fields with shift modulation for multi-sequence MRI translation}

\author{\firstname Yunjie \surname Chen \orcid{0000-0001-9478-6953} \email y.chen@lumc.nl \\  
	\addr Department of Radiology, Leiden University Medical Center, Leiden, the Netherlands
	\AND
        \firstname Marius \surname Staring \orcid{0000-0003-2885-5812} \email m.staring@lumc.nl \\
        \addr Department of Radiology, Leiden University Medical Center, Leiden, the Netherlands 
        \AND
        \firstname Olaf M. \surname Neve \orcid{0000-0002-5104-8448} \email o.m.neve@lumc.nl \\
        \addr Department of Otorhinolaryngology and Head \& Neck Surgery, Leiden University Medical Center, \\ Leiden, the Netherlands
        \AND
        \firstname Stephan R. \surname Romeijn \orcid{0000-0002-4634-447X} \email s.r.romeijn@lumc.nl \\
        \addr Department of Radiology, Leiden University Medical Center, Leiden, the Netherlands 
        \AND
        \firstname Erik F. \surname Hensen \orcid{0000-0002-4393-7421} \email e.f.hensen@lumc.nl \\
        \addr Department of Otorhinolaryngology and Head \& Neck Surgery, Leiden University Medical Center, \\ Leiden, the Netherlands 
        \AND
        \firstname Berit M. \surname Verbist \orcid{0000-0002-1010-2583} \email b.m.verbist@lumc.nl \\
        \addr  Department of Radiology, Leiden University Medical Center, Leiden, the Netherlands 
        \AND
        \firstname Jelmer M. \surname Wolterink \orcid{0000-0001-5505-475X} \email j.m.wolterink@utwente.nl \\
        \addr Department of Applied Mathematics, Technical Medical Center, University of Twente, Enschede, the Netherlands 
        \AND
        \firstname Qian \surname Tao \orcid{0000-0001-7480-0703} \email q.tao@tudelft.nl \\
        \addr Department of Imaging Physics, Delft University of Technology, Delft, the Netherlands
}

\begin{document}

\maketitle

\begin{abstract}
    Multi-sequence magnetic resonance imaging (MRI) has found wide applications in both modern clinical studies and deep learning research. However, in clinical practice, it frequently occurs that one or more of the MRI sequences are missing due to different image acquisition protocols or contrast agent contraindications of patients, limiting the utilization of deep learning models trained on multi-sequence data. One promising approach is to leverage generative models to synthesize the missing sequences, which can serve as a surrogate acquisition. State-of-the-art methods tackling this problem are based on convolutional neural networks (CNN) which usually suffer from spectral biases, resulting in poor reconstruction of high-frequency fine details. In this paper, we propose Conditional Neural fields with Shift modulation (CoNeS), a model that takes voxel coordinates as input and learns a representation of the target images for multi-sequence MRI translation. The proposed model uses a multi-layer perceptron (MLP) instead of a CNN as the decoder for pixel-to-pixel mapping. Hence, each target image is represented as a neural field that is conditioned on the source image via shift modulation with a learned latent code. Experiments on BraTS 2018 and an in-house clinical dataset of vestibular schwannoma patients showed that the proposed method outperformed state-of-the-art methods for multi-sequence MRI translation both visually and quantitatively. Moreover, we conducted spectral analysis, showing that CoNeS was able to overcome the spectral bias issue common in conventional CNN models. To further evaluate the usage of synthesized images in clinical downstream tasks, we tested a segmentation network using the synthesized images at inference. The results showed that CoNeS improved the segmentation performance when some MRI sequences were missing and outperformed other synthesis models. We concluded that neural fields are a promising technique for multi-sequence MRI translation.
    Our code is available at~\url{https://github.com/cyjdswx/CoNeS.git}.
\end{abstract}


\section{Introduction}
    Multi-sequence magnetic resonance imaging (MRI) plays a key role in radiology and medical image computing. One advantage of MRI is the availability of various pulse sequences, such as T1-weighted MRI (T1), T2-weighted MRI (T2), T1-weighted with contrast (T1ce), and T2-fluid-attenuated inversion recovery MRI (FLAIR), which can provide complementary information to clinicians \citep{Cherubini2016}. The importance of the availability of multi-sequence MRI was also indicated by recent deep learning research \citep{cercignani2018}, which shows that the more sequences were used for segmentation, the better results could be obtained. However, due to clinical restrictions on the use of contrast agents and the diversity in imaging protocols in different medical centers, it is difficult and time-consuming to always obtain exactly the same MRI sequences for training and inference, which may damage the generalization and performance of deep learning segmentation models. 
    
    One way to tackle this problem is to generate missing sequences from existing images based on the information learned from a set of paired images, known as image-to-image translation. Like in other computer vision tasks, convolutional neural networks (CNNs) with an encoder and decoder architecture are normally used for this specific task \citep{sevetlidis2016,joyce2017,wei2019fluid}. Despite the significant improvement over traditional non-deep-learning methods, these methods still suffer from the limitation of a pixel-wise loss function, such as the L1 or MSE loss, which tends to result in blurry results with undesirable loss of details in image structures \citep{isola2017,dalmaz2022resvit}. To overcome this limitation, generative adversarial networks (GANs) were introduced for image-to-image translation and rapidly became a training protocol benchmark for medical image translation \citep{li2019diamondgan,nie2018medical,armanious2020medgan}. GANs improve translation results both visually and quantitatively owing to the adversarial learning loss, which penalizes the images that are correctly classified by the discriminator. 
    
    However, research showed that generative models that use a CNN as a backbone network consisting of ReLU activation functions and transposed or up-convolutional layers usually suffer from spectral biases \citep{rahaman2019spectral,durall2020watch}. Therefore, these generative models fit low-frequency signals first and may again fail to capture details in image structures during training. Transformers, which instead use multi-head self-attention blocks and multi-layer perceptrons (MLPs) have gained tremendous attention in computer vision research \citep{liu2023cascaded, jiang2021transgan}. Due to the absence of convolutional layers, transformers show great potential for preserving fine details and long-range dependencies and have recently been applied to medical image translation \citep{liu2023one,dalmaz2022resvit}. However, despite the numerous efforts made by these studies, such as hybrid architectures and image patch-based processing, the training of transformers is still considered heavy and data-demanding \citep{dosovitskiy2020image,esser2021taming}. The inherently high computational complexity of the transformer block and expensive memory cost of low-level tasks, such as denoising and super-resolution, further complicate the application of transformers in medical image translation \citep{chen2021pre}. 

    To address these limitations, we propose image-to-image translation using neural fields \citep{xie2022neural}. In contrast to CNN or transformer-based methods, a neural field represents the target images on a continuous domain using a coordinate-based network, which can be conditioned on the information extracted from the source images. We previously proposed an image-to-image translation approach based on neural fields \citep{chen2023}. Here, we substantially extend this model by proposing \textbf{Co}nditional \textbf{Ne}ural fields with \textbf{S}hift modulation (CoNeS). In contrast to traditional deep learning computer vision techniques, CoNeS parameterizes the target images as neural fields that can be queried on a grid to provide pixel-wise predictions. Specifically, we use an MLP as the decoder to map the voxel coordinates to the intensities on the target images. To capture instance-specific information, we condition the neural fields on the latent codes extracted from the source images. By applying shift modulation, the neural fields can be further varied across the coordinates to enhance their ability to preserve high-frequency signals.
    
    Although plenty of work has shown great progress in medical image translation, most previous works have been evaluated based on image similarity metrics and only a few papers have evaluated the benefits of using synthesized images for downstream analysis. \cite{amirrajab2023} fine-tuned a segmentation model with synthesized cardiac images to improve the performance of different modalities; \cite{skandarani2020effectiveness} introduced a variational auto-encoder (VAE) based network for image translation based data augmentation to improve the generalization capabilities of a segmentation model. In practice, however, it would be more straightforward and beneficial to use the synthesized images directly without fine-tuning or training a new network. In this study, we perform downstream experiments using a pre-trained segmentation model to further evaluate different image translation models. 
    
    The main contributions of our work are:
    \begin{itemize}
        \item We developed a novel generative adversarial network for medical image translation based on conditional neural fields. In the proposed model, we build neural fields on the coordinates across the image to fit the target image. To improve the performance and the stability of the model, we introduce shift modulation, which conditions the neural fields on the output of a hypernetwork.

        \item We evaluated the proposed model by synthesizing various MRI sequences on two paired multi-sequence brain MRI datasets. The results show that the proposed model outperforms state-of-the-art methods both visually and quantitatively. We additionally performed spectral analysis, which indicates that our method is not affected by spectral biases in the way that traditional CNN-based generative models are.
        
        \item We compare different medical image translation models in downstream tasks by testing a segmentation model with the synthesized images. Our experiments indicate that by applying image translation, we can improve segmentation performance for incomplete MRI acquisition and our synthesized images outperform the state-of-the-art methods.
    \end{itemize}

\section{Related work}
\label{rec:related_work}
    \paragraph{Missing MRI sequences} Several studies have dealt with the missing MRI sequences problem in medical image analysis \citep{Azad2022}. One early idea was to translate all available sequences into a shared latent space for downstream analysis. Following this idea, \cite{Havaei2016} developed the Hetero-Modal Image Segmentation (HeMIS) method, where sequence-specific convolutional layers are applied to each image sequence for establishing a common representation which enables robust segmentation when some images are missing. Later, \cite{hu2020} and \cite{azad2022smu} introduced knowledge distillation to improve the segmentation performance in the same situation. In such a model, a network using all modalities as input (teacher network) and another network using a subset of them (student network) are optimized synchronously. During training, information from all modalities is distilled from the teacher to the student network to improve the performance of the student network. Recently, \cite{liu2023cascaded} developed a transformer-based model for Alzheimer's classification that can handle missing data scenarios. All these models managed to build a robust model for the situation that only a part of the modalities are available. However, since the missing MRIs are not explicitly constructed, it is still difficult for medical doctors to interpret and make decisions with these methods in clinical practice.
    
    \paragraph{Image-to-image translation} Image-to-image translation, on the contrary, focuses on synthesizing missing images from existing ones based on prior knowledge learned from the dataset. By predicting the missing images, clinicians can offer comprehensive diagnoses and also find an explanation of the results in downstream analysis. Recent progress in generative modeling, such as generative adversarial networks (GANs), variational auto-encoders (VAE), and diffusion models, has shown extraordinary capabilities in image generation \citep{isola2017,kawar2022}. In the domain of medical image translation, \cite{dar2019image} proposed pGAN based on a conditional GAN combined with a pixel-wise loss and a perceptual loss. \cite{sharma2019missing} proposed a multi-modal GAN (MM-GAN) that extends the idea by using multi-modality imputation for arbitrary input and output modalities. Recently, \cite{yurt2021mustgan} proposed mustGAN that enhanced the synthesis performance by aggregating multiple translation streams. Inspired by the recent progress of the transformer model, \cite{dalmaz2022resvit} proposed ResViT based on a hybrid architecture that consists of convolutional operators and transformer blocks. Although promising, most studies focus on the image quality of the output images, and only a few have extended their work to the use of synthesized images in downstream tasks \citep{iglesias2013synthesizing,van2015does,amirrajab2023}.
    
    \paragraph{Neural fields} Neural fields, also known as implicit neural representations (INRs) or coordinate-based networks, are increasingly popular in computer vision and medical image analysis \citep{xie2022neural,molaei2023implicit}. The core idea behind neural fields is that neural networks are not used to learn an operator between signals, as in CNNs or vision transformers, but to \textit{represent} a complex signal on a continuous spatial or spatiotemporal domain. Neural fields can be used to solve a wide range of problems, including 3D scene reconstruction and generative shape modeling. \cite{park2019deepsdf} proposed DeepSDF which learns a continuous signed distance function to represent 3D surfaces. One distinguished benefit of using neural fields is the capability to handle data with variable resolution because of the absence of up-sampling architectures. Inspired by this, \cite{chen2021learning} proposed a Local Implicit Image Function (LIIF) for image super-resolution, which also shows the potential of handling image generation. Similarly in the field of medical imaging, \cite{mcginnis2023single} performed multi-contrast MRI super-resolution via neural fields without any high-resolution training data. \cite{wolterink2022implicit} proposed to use INRs to represent a transformation function for deformable image registration. \cite{amiranashvili2022learning} proposed to reconstruct anatomical shapes from sparse measurements via neural fields. Recently, \cite{Shaham2021} developed Spatially-Adaptive Pixelwise Networks (ASAP-Net), which is most relevant to our work, to speed up image-to-image translation by locally conditioned MLPs. Different from prior work, the neural fields in CoNeS are conditioned on a latent code varying across the coordinates through shift modulation, inspired by \cite{dupont2022data}. Specifically, CoNeS consists of a global MLP shared by the whole image and a varying latent code, which determines pixel-wise affine transformations to modulate the neural fields. 

    \begin{figure}[tb]
	\centering
	\includegraphics[width=\linewidth]{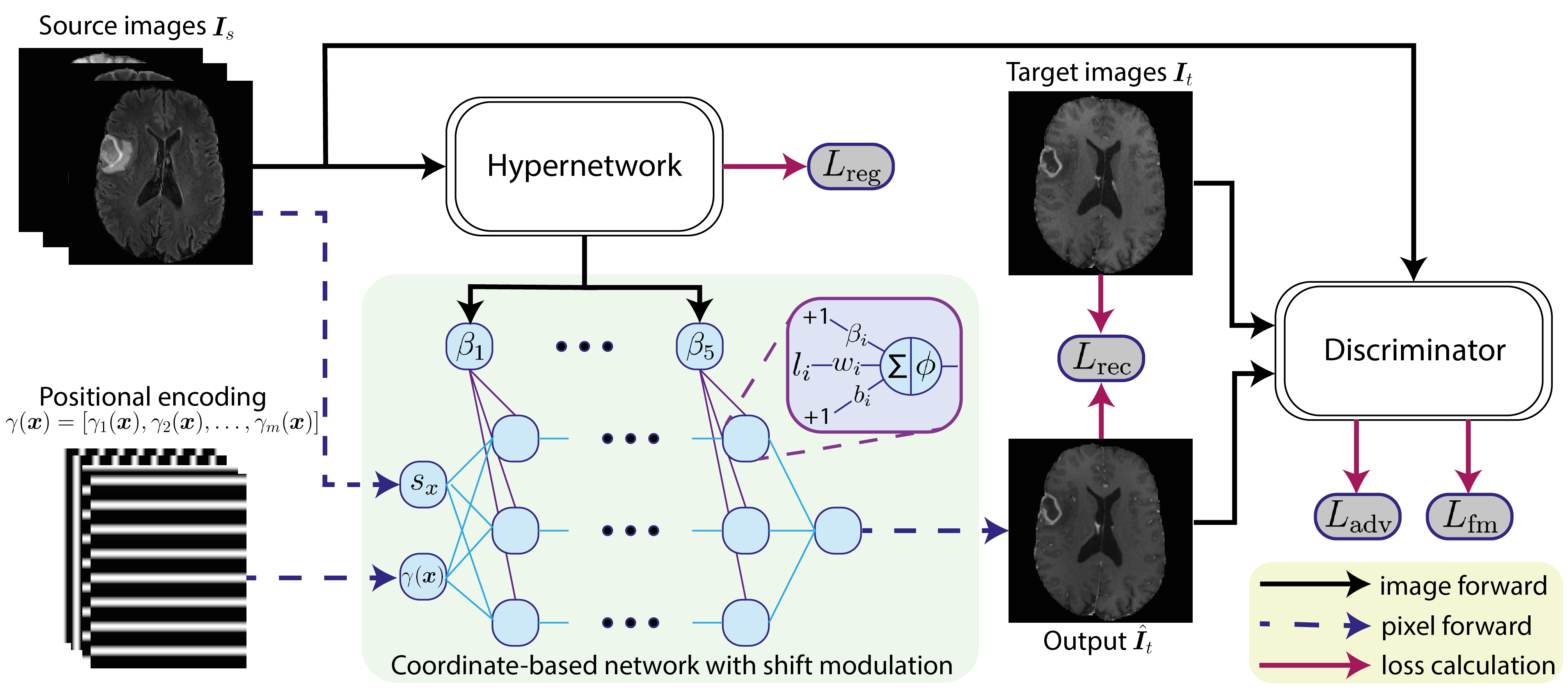}
	\caption{The overall architecture of CoNeS. The generator in the proposed models consists of a hypernetwork and a coordinate-based network. We condition the coordinate-based network on a varying latent code, which is generated by the hypernetwork, across coordinates via shift modulation. The conditional discriminator, which takes both the source images and real/fake images as input, further improves the performance of the generator. The proposed model is optimized using a reconstruction loss $L_{\text{rec}}$, an adversarial loss $ L_{\text{adv}}$, a feature matching loss $L_{\text{fm}}$ and latent code regularization $L_{\text{reg}}$.}
        \label{fig:overallachitecture}
    \end{figure}
 
\section{Methods}
    \subsection{Model overview}
        To formulate the problem, let $\bm{I}_t =\{I_t^i\}^{N_t}_{i=0}$ be the set of missing MRI sequences and $\bm{I}_s =\{I_s^i\}^{N_s}_{i=0}$ be the set of available MRI sequences, where $N_t$ is the number of target sequences and $N_s$ the number of source sequences. We assume that all images from an instance are co-registered so that there is no extra deformation between the images. As a result, our problem is identical to learning a mapping function $\Phi: \bm{I}_s \rightarrow \bm{I}_t$ using the training dataset, which can be applied to all patients in the test dataset and generate the corresponding missing image $\bm{I}_t$. Similar to traditional GAN models, the proposed model consists of a generator that performs the mapping and a discriminator that aims to tell the real target image and the synthesized one apart. As introduced in pix2pix \citep{isola2017}, we apply a conditional discriminator that takes both the source and predicted image as its input. The overall architecture of our approach is shown in Fig.~\ref{fig:overallachitecture}. In the following section, we introduce how to use a coordinate-based network to model conditional neural fields for image-to-image translation. 

    \subsection{Coordinate-based network}  
        In a typical neural field algorithm, a quantity defined over spacetime, such as an RGB intensity or a signed distance function, is represented as a neural network that maps coordinates to the quantity. Specifically in our problem, we train an MLP that takes coordinates as input and outputs intensities of the target MRI sequences. Given a normalized $d$-dimensional coordinate $\bm{x} \in \mathbb{R}^d$, where each component lies in [-1,1], we use $\bm{t}_x = \{t^i_{\bm{x}}\}$ and $\bm{s}_x = \{s^i_{\bm{x}}\}$ to denote the intensities at position $\bm{x}$, where $t_x^i$ refers to the intensity value in $I_t^i$ and $s^i_x$ refers to the intensity value in $I_s^i$, respectively. Hence, the function $\Phi$ can be formulated as a pixel-wise mapping that generates intensities over a d-dimensional space:
        \begin{linenomath}
            \begin{align}
                \bm{t}_{\bm{x}} &= \Phi(\bm{x};\bm{z}),
                \label{eq:mapping}
            \end{align}
        \end{linenomath}
        where $\bm{z}$ is a latent code that contains instance-specific information. During inference, the target images $\hat{\bm{I}_t} = \{\hat{I^i_t}\}$ are obtained by intensity prediction via sampling from the entire grid via $\Phi$.
        
        A network directly operating on the Cartesian coordinates tends to fit the low-frequency signals first and, as a result, fails to reconstruct the high-frequency image details \citep{mildenhall2021,rahaman2019spectral}. One popular approach to overcome this problem is to map the Cartesian coordinates to a higher dimensional space via positional encoding $\gamma: \mathbb{R}^d \rightarrow \mathbb{R}^m$. In the proposed model, we use sinusoidal functions to perform positional encoding as follows \citep{zhong2019reconstructing}:
        \begin{linenomath}
            \begin{align}
            \gamma(\bm{x}) &= [\gamma_1(\bm{x}), \gamma_2(\bm{x}), \ldots, \gamma_m(\bm{x})],\\
            \gamma_{2i}(\bm{x}) &= \sin(2^{i-1}\pi \bm{x}),\\
            \gamma_{2i+1}(\bm{x}) &= \cos(2^{i-1}\pi \bm{x}),
            \label{eq:positionalencoding}
            \end{align}
        \end{linenomath}
        where $m$ is a frequency parameter. Positional encoding can also be seen as Fourier feature mapping of the Cartesian coordinates. By using positional encoding as the input of the MLP, we enable the network to fit the neural field containing high-frequency variation. 
    
    \subsection{Conditional neural fields}
        To let the neural field adapt to different input images, we condition it on a set of latent codes $\bm{z}$, which contain instance-specific information. In the proposed model, we introduce a hypernetwork $H$ that generates the latent code from the source images: $\bm{z}=H(\bm{I}_s)$. By extracting $\bm{z}$, we can then vary and adapt the neural fields to different instances. Below, we explain how we obtain the latent code $\bm{z}$ and how the proposed method parameterizes the neural fields with the conditioning via $\bm{z}$.

        \subsubsection{Hypernetwork}
        A hypernetwork refers to an extra neural network that generates parameters for the main network \citep{ha2017hypernetworks}. The main network behaves like a typical neural network, while the hypernetwork encodes information from the inputs and transfers the information to the main network via the generated parameters. For clarity, we use $\bm{z}_i = [\bm{\alpha}_i, \bm{\beta}_i]$ to denote the latent code used by the i-th layer of the MLP, where $\bm{\alpha}_i$ are the weights and $\bm{\beta}_i$ are the biases, both generated by $H$. Hence, for each layer of the MLP, we have:
        \begin{linenomath}
            \begin{align}
            \bm{l}_{i+1} = \phi(\bm{\alpha}_i\bm{l}_i + \bm{\beta}_i),
            \label{eq:MLP}
            \end{align}
        \end{linenomath}
        where $\bm{l}_i$ is the input feature of the i-th layer, and $\phi$ is the activation function. Both $\bm{\beta}_i$ and $\bm{l}_i$ are column vectors of size $n_{i+1} \times 1$ and $\bm{\alpha}_i$ is a matrix of size $n_{i+1} \times n_{i}$, where $n_{i}$ is the number of neurons of the i-th layer. Inspired by ASAP-Net \citep{Shaham2021}, we vary the neural field of each pixel by varying the latent code across the coordinates, which can be denoted as $\bm{z}_i(\bm{x}) = [\bm{\alpha}_i(\bm{x}), \bm{\beta}_i(\bm{x})]$, to improve the representation capability. We use $H_{\bm{x}}$ to represent the latent code mapping for each pixel, and thus, $\Phi$ can be denoted as:
        \begin{linenomath}
            \begin{align}
            t_{\bm{x}} = \Phi(\gamma(\bm{x});\bm{z}(\bm{x})) = \Phi(\gamma(\bm{x});H_{\bm{x}}(\bm{I}_s)),
            \label{eq:mappingwithhyper}
            \end{align}
        \end{linenomath}
        and each layer of the MLP can be denoted as:
        \begin{linenomath}
            \begin{align}
            \bm{l}_{i+1}(\bm{x}) = \phi(\bm{\alpha}_i(\bm{x})\bm{l}_i(\bm{x}) + \bm{\beta}_i(\bm{x})), 
            \label{eq:MLP_pixelwise}
            \end{align}
        \end{linenomath}
        where $\bm{l}_i(\bm{x})$ refers to the i-th input feature at position $\bm{x}$. Different from ASAP-Net, we adapt the bottom-up pathway from a feature pyramid network \citep{lin2017feature} as the hypernetwork $H$, which outputs the latent code $\bm{z}(\bm{x})$ for each pixel with a feasible memory cost (detailed in Section~\ref{sec:implementation}).

        \subsubsection{Shift modulation}
        By conditioning neural fields on varying latent codes across the coordinates, we can improve the representation capability of the network and better model the structure details \citep{xie2022neural,peng2020convolutional}. However, the number of parameters also increases with the spatial expansion of the neural fields, which may induce high computational costs and damage the performance due to over-fitting. This problem may become worse with larger input images. To compact the model while maintaining spatially varying neural fields, we propose to condition the neural network through feature-wise linear modulation (FiLM) \citep{perez2018film}. Instead of generating all parameters of the MLP per pixel, an affine transformation (scale and shift) is applied to every neuron of a single, global MLP. Thus, each layer of the one MLP can be denoted as:
        \begin{linenomath}
            \begin{gather}
            \bm{l}_{i+1}(\bm{x}) = \overline{\bm{\alpha}_i}(\bm{x})\phi(\bm{w}_i\bm{l}_i + \bm{b}_i) +\bm{\beta}_i(\bm{x}),
            \label{eq:FILM}
            \end{gather}
        \end{linenomath}
        where the weights and biases of the MLP are now replaced by trainable parameters $\bm{w}_i$ and $\bm{b}_i$ that are shared by all coordinates, and $\overline{\bm{\alpha}_i}(\bm{x})$ is an $n_{i+1} \times n_{i+1}$ matrix that performs scaling to the neurons of the i-th layer by left matrix multiplication. Thus, we can obtain a modified neural field for each coordinate with fewer parameters. Furthermore, research shows that by using shifts only, which is so-called shift modulation, we can achieve comparable results with half of the parameters \citep{dupont2022data}. In this case, $\overline{\bm{\alpha}_i}$ is an identity matrix and all latent codes are used as shift parameters: $\bm{z}_i(\bm{x}) = \bm{\beta}_i(\bm{x})$. In practice, we split the biases of the MLP into two parts: trainable biases $\bm{b}_i$ and biases $\bm{\beta}_i(\bm{x})$ generated by $H$:
        \begin{linenomath}
            \begin{gather}
            \bm{l}_{i+1}(\bm{x}) = \phi(\bm{w}_i\bm{l}_i + \bm{b}_i + \bm{\beta}_i(\bm{x})).
            \label{eq:shiftmodulation}
            \end{gather}
        \end{linenomath}
        The hypernetwork $H$ is optimized together with the MLP during training. In the experimental section, we will show by using shift modulation our model can achieve better performance at reduced complexity.

        \subsubsection{Intensity concatenation} 
        In addition to shift modulation, we also condition the neural fields on the source images directly. Different from the latent codes, the pixel intensities provide first-hand uncoded local information. We concatenate the image intensities from all the source images as an additional input of the MLP. The mapping function of the neural fields therefore becomes:
        \begin{linenomath}
            \begin{gather}
                t_{\bm{x}} = \Phi(\gamma(\bm{x}),\bm{s}_{\bm{x}};H_{\bm{x}}(\bm{I}_s)).
                \label{eq:finalmappingfunction}
            \end{gather}
        \end{linenomath}   
        
        \subsection{Loss function}
        Like the standard GAN model, the discriminator and the generator in the proposed model are optimized alternately. In each iteration, we train the discriminator using the hinge loss \citep{lim2017geometric}:
        \begin{linenomath}
            \begin{align}
            L_D = \mathbb{E}_{\bm{I}_t,\bm{I}_s}[\max(0,1-D(\bm{I}_t,\bm{I}_s))] + \mathbb{E}_{\bm{I}_s}[\max(0,1+D(\hat{\bm{I}_t},\bm{I}_s))],
            \end{align}
        \end{linenomath}
        where $D$ is the discriminator and $\mathbb{E}$ is the expectation over the whole dataset.
        
        The generator is trained by a loss function $L$ that contains a reconstruction loss, an adversarial loss, a feature matching loss, and latent code regularization.
        
        \paragraph{Reconstruction loss} To ensure the synthesized images are as close to the real images as possible, we apply a reconstruction loss that maximizes the similarity between ground truth $\bm{I}_t$ and output images $\hat{\bm{I}_t}$, which are obtained by intensity prediction via sampling from the entire grid. We use the $\ell_1$ loss function as suggested in \citep{isola2017}:
        \begin{linenomath}
            \begin{align}
            L_{\text{rec}} = \mathbb{E}_{\bm{I}_t,\bm{I}_s}[\lVert \hat{\bm{I}_t}-\bm{I}_t \rVert_1].
                \label{eq:recloss}
            \end{align}
        \end{linenomath}
        \paragraph{Adversarial loss} Adversarial loss is applied to enforce that the generated images are good enough to fool the discriminator. Like the discriminator loss, we use the hinge function, which is defined as:
        \begin{linenomath}
            \begin{align}
            L_{\text{adv}} &= -\mathbb{E}_{\bm{I}_s}[\log D(\hat{\bm{I}_t},\bm{I}_s)].
            \label{eq:advloss}
            \end{align}
        \end{linenomath}
        \paragraph{Feature matching loss} To stabilize the training, we apply a feature matching loss introduced by \cite{wang2018high}. Specifically, we feed both the real and generated images to the discriminator and extract the intermediate features from each forward pass. The two groups of the intermediate features are matched using the $\ell_1$ loss function. Hence, the feature matching loss is defined as:
        \begin{linenomath}
            \begin{align}
            L_{\text{fm}} = \mathbb{E}_{\bm{I}_t,\bm{I}_s}\sum_{i=1}^{T}\frac{1}{N_i}[\lVert D_{i}(\bm{I}_t,\bm{I}_s) - D_{i}(\hat{\bm{I}_t},\bm{I}_s) \rVert_1].
            \label{eq:fmloss}
            \end{align}
        \end{linenomath}
        where $D_{i}$ denotes the i-th feature layer of the discriminator and $N_i$ denotes the number of elements in each layer. $T$ is the total number of layers of the discriminator.
        \paragraph{Latent code regularization} Last, we apply the $\ell_2$ norm to $\bm{z}$ as a latent code regularization to stabilize the training:
        \begin{linenomath}
            \begin{align}
            L_{\text{reg}} = \mathbb{E}_{\bm{I}_s}\lVert H_{\bm{x}}(\bm{I}_s)\rVert_2.
            \label{eq:latentreg}
            \end{align}
        \end{linenomath}
        \paragraph{Overall loss} The overall loss function then becomes
        \begin{linenomath}
            \begin{align}
            L = \lambda_{\text{rec}}L_{\text{rec}} + \lambda_{\text{adv}}L_{\text{adv}} + \lambda_{\text{fm}}L_{\text{fm}}+\lambda_{\text{reg}}L_{\text{reg}},
            \label{eq:totalloss}
            \end{align}
        \end{linenomath}
        where $\lambda_{\text{rec}}$, $\lambda_{\text{adv}}$,
        $\lambda_{\text{fm}}$, and $\lambda_{\text{reg}}$ are the weights of the loss functions.
    
\section{Experiments and results}

    \subsection{Dataset}
    To evaluate the proposed translation model, we conducted experiments on two datasets: (1) BraTS 2018 \citep{menze2014multimodal} and (2) an in-house Vestibular Schwannoma MRI (VS) dataset \citep{neve2022fully}. 
    
    \paragraph{BraTS 2018} BraTS 2018 is a multi-sequence brain MRI dataset for tumor segmentation. The dataset consists of 285 patients for training and 66 patients for validation. Each patient has four co-registered MRI sequences: T1 (1-6 mm slice thickness), T1ce (1-6 mm slice thickness), T2 (2-6 mm slice thickness) and FLAIR (2-6 mm slice thickness). The tumor mask that includes the non-enhanced tumor, the enhanced tumor, and the edema was delineated by experts from multiple centers as a segmentation ground truth. All the scans in BraTS 2018 are resampled to 1 mm isotropic resolution.
    
    \paragraph{Vestibular schwannoma MRI dataset} The VS dataset is MRI scans of patients with vestibular schwannoma, which is a benign tumor arising from the neurilemma of the vestibular nerve. 191 patients were collected from 37 different hospitals using 12 different MRI scanners. In our study, 147 patients are selected for training and the remaining 44 patients are the validation set. All patients have a gadolinium-enhanced T1-weighted MRI (shortened to T1ce) and a high-resolution T2 (shortened to T2). The spatial resolution of the T1ce ranges from $0.27 \times 0.27 \times 0.9$ to $1.0 \times 1.0 \times 5.0$ mm, and the spatial resolution of T2 scans ranges from $0.23 \times 0.23 \times 0.5$ to $0.7 \times 0.7 \times 1.8$ mm. The intra- and extrameatal tumor was manually delineated by four radiologists. Different from BraTS 2018, there are only two sequences available in the VS dataset and the high resolution of the T2 offers better visibility of the lesion but may also degrade the image quality, which makes the image translation more challenging on this dataset.  
    
    \subsection{Experimental setup}
    \label{sec:implementation}
     \paragraph{Network architecture} The hypernetwork $H$ of the proposed model is adapted from feature pyramid network \citep{lin2017feature}. $H$ consists of four convolutional modules containing [2, 4, 23, 3] residual blocks in sequence. Each convolutional module is followed by a $3 \times 3$ convolutional smoothing layer and up-sampling layer, computing a feature map that is downscaled by a factor of 2. We take the output of the last module, which has the same size as the input resolution, as the latent code $\bm{z}$. Adapted from  \citep{Shaham2021}, the MLP in the proposed model contains five 64-channel layers. The Leaky ReLU function with a negative slope of 0.2 is applied as the activation function after all intermediate layers. The output layer is followed by a Tanh function which can constrain the range of the intensities to [-1, 1]. The discriminator is a 2D convolutional neural network that takes both the source image and prediction as input, both as a whole. The network contains five $4 \times 4$ convolutional blocks followed by a Leaky ReLU function except for the last layer. The strides and number of filters of the blocks are [2, 2, 2, 1, 1] and [64, 128, 256, 512, 1] respectively. Like pix2pix \citep{isola2017}, the discriminator down-samples the inputs by 8 and penalizes structures at the scale of patches.
    
     \paragraph{Pre-processing} Registration was applied to the VS dataset before training. We considered the T1ce as the fixed image and performed rigid registration with the T2, for which we used Elastix software \citep{klein2009elastix}. All images from both datasets were then normalized to the range of [-1,1] and the background was cropped based on the bounding box of the foreground before training to reduce the image size. Both sequences in the VS dataset were resampled to $0.29\times 0.29$ mm in-plane resolution, which is the median value of the T1ce domain. During training, random cropping was conducted on the images, with a cropping size of $160 \times 128$ for BraTS 2018 and a cropping size of $320 \times 320$ for the VS dataset, respectively. 
     
     \paragraph{Implementation details} All experiments were conducted using Python 3.10 and PyTorch 1.12.1 on a mixed computation server equipped with Nvidia Quadro RTX 6000 and Nvidia Tesla V100 GPUs. The models were trained by the Adam optimizer using the Two Time-scale Update Rule (TTUR) training scheme, in which the generator and discriminator have a different initial learning rate \citep{heusel2017gans}. We found that an initial learning rate of $1 \times 10^{-4}$ for the generator and an initial learning rate of $4 \times 10^{-4}$ for the discriminator worked best for our experiments. The learning rates were further decayed using a linear learning rate scheduler. Adapted from the choices of hyperparameters in \cite{Shaham2021}, we set the frequency parameter $m=6$ for positional encoding. We set $\lambda_{\text{adv}} = 1.0$ and $\lambda_{\text{rec}} = 100.0$, which gives us the best balance between sharp results and fewer artifacts as suggested in \cite{isola2017}. Both $L_{\text{fm}}$ and $L_{\text{reg}}$ help to stabilize the training, while large $\lambda_{\text{reg}}$ and $\lambda_{\text{fm}}$ may lead to poor reconstruction performance. We set $\lambda_{\text{fm}}=\lambda_{\text{reg}}=10.0$, which ensures a stable training while maintaining reconstruction performance \citep{wang2018high, Shaham2021}. Lastly, we focus on 2D image translation in this paper and hence use 2D coordinates ($d=2$).

    \begin{table}[tb] 
		\centering
        \small
		\caption{Quantitative comparison of different image translation models on BraTS 2018. The mean value and standard deviation of PSNR and SSIM are reported. The highest values per column are indicated in boldface; The $\dagger$ after each metric of the benchmarks indicates a significant difference (\(p < .05\)) compared to the proposed method.}
		\begin{tabular}{lcccccccc}
		\multirow{2}{*}{\textbf{model}} &  \multicolumn{2}{c}{\textbf{T1ce translation}} & \multicolumn{2}{c}{\textbf{T1 translation}} & \multicolumn{2}{c}{\textbf{T2 translation}} & \multicolumn{2}{c}{\textbf{FLAIR translation}} \\
			\cline{2-9} 
            & PSNR & SSIM & PSNR  & SSIM & PSNR  & SSIM & PSNR  & SSIM\\
            \hline
		\multirow{2}{*}{pix2pix} & 30.1 & 0.941 & 27.0 & 0.945 & 28.0 & 0.926 & 27.6 & 0.910\\
            & $\pm 2.65^{\dagger}$ & $\pm 0.014^{\dagger}$ & $\pm 3.69$ & $\pm 0.013^{\dagger}$ & $\pm 2.63^{\dagger}$  & $\pm 0.062^{\dagger}$ & $\pm 3.03^{\dagger}$ & $\pm 0.097^{\dagger}$\\
            \hline
            \multirow{2}{*}{pGAN} & 30.7 & 0.943 & $\bm{27.5}$ & 0.945 & 29.2  & 0.943 & 28.5 & 0.916\\
            & $\pm 3.18^{\dagger}$ & $\pm 0.015^{\dagger}$ & $\bm{\pm 3.72}$ & $\pm 0.015^{\dagger}$ & $\pm 2.80^{\dagger}$ &  $\pm 0.020^{\dagger}$ & $\pm 3.26^{\dagger}$ & $\pm 0.095^{\dagger}$\\
            \hline
            \multirow{2}{*}{ResViT} & 29.2 & 0.935 & 25.0 & 0.918 &  26.6 & 0.923 & 24.7 & 0.876\\
            & $\pm 2.37^{\dagger}$ & $\pm 0.014^{\dagger}$ & $\pm 2.60^{\dagger}$ & $\pm 0.014^{\dagger}$ &  $\pm 2.30^{\dagger}$ & $\pm 0.020^{\dagger}$ & $\pm 2.08^{\dagger}$ & $\pm 0.092^{\dagger}$\\
            \hline
            \multirow{2}{*}{ASAP-Net} & 30.8 & 0.948 & 27.3 & 0.948 & 28.6 & 0.940 & 28.4 & 0.916\\
            & $\pm 2.97^{\dagger}$ & $\pm 0.017^{\dagger}$ & $\pm 3.79$ & $\pm 0.015^{\dagger}$ & $\pm 2.74^{\dagger}$ & $\pm 0.019^{\dagger}$ & $\pm 3.10^{\dagger}$ & $\pm 0.098^{\dagger}$\\
            \hline
			CoNeS  & $\bm{31.2}$ & $\bm{0.951}$ & 27.3 & $\bm{0.953}$ & $\bm{29.6}$ & $\bm{0.950}$ & $\bm{29.1}$ & $\bm{0.926}$ \\
                (proposed) & $\bm{\pm 3.11}$ & $\bm{\pm 0.017}$ & $\pm 4.03$ & $\bm{\pm 0.014}$ & $\bm{\pm 3.03}$ & $\bm{\pm 0.021}$ & $\bm{\pm 2.99}$ & $\bm{\pm 0.097}$
		\end{tabular}
            \label{tb:bratsimagequality}
    \end{table}

    \begin{table}[tb] 
		\centering
        \small
		\caption{Quantitative comparison of different image translation models after cropping on BraTS 2018. The mean value and standard deviation of PSNR and SSIM are reported. The highest values per column are indicated in boldface; The $\dagger$ after each metric of the benchmarks indicates a significant difference (\(p < .05\)) compared to the proposed method.}
		\begin{tabular}{lcccccccc}
		\multirow{2}{*}{\textbf{model}} &  \multicolumn{2}{c}{\textbf{T1ce translation}} & \multicolumn{2}{c}{\textbf{T1 translation}} & \multicolumn{2}{c}{\textbf{T2 translation}} & \multicolumn{2}{c}{\textbf{FLAIR translation}} \\
			\cline{2-9} 
            & PSNR & SSIM & PSNR & SSIM & PSNR & SSIM & PSNR & SSIM\\
            \hline
			\multirow{2}{*}{pix2pix} & 19.9 & 0.610 & 15.7 & 0.636 & 19.9  & 0.658 & 18.1 & 0.585\\
            & $\pm 3.30^{\dagger}$ & $\pm 0.092^{\dagger}$ & $\pm 4.41$ & $\pm 0.089^{\dagger}$ & $\pm 2.85^{\dagger}$  & $\pm 0.082^{\dagger}$ & $\pm 3.31^{\dagger}$ & $\pm 0.081^{\dagger}$\\
            \hline
            \multirow{2}{*}{pGAN} & 20.6 & 0.646 & $\bm{16.1}$ & 0.663 & 20.8 & 0.721 & 18.9 & 0.636\\
            & $\pm 3.73^{\dagger}$ & $\pm 0.096^{\dagger}$ &  $\bm{\pm 4.61}$ & $\pm 0.099$ & $\pm 3.23^{\dagger}$ & $\pm 0.093^{\dagger}$ & $\pm 3.70^{\dagger}$ & $\pm 0.086^{\dagger}$\\
            \hline
            \multirow{2}{*}{ResViT} & 20.2 & 0.612 & 15.1 & 0.599 & 19.5 & 0.672 & 17.0 & 0.545\\
            & $\pm 3.56^{\dagger}$ & $\pm 0.098^{\dagger}$ & $\pm 4.32^{\dagger}$ & $\pm 0.090^{\dagger}$ & $\pm 3.50^{\dagger}$ & $\pm 0.093^{\dagger}$ & $\pm 3.26^{\dagger}$ & $\pm 0.111^{\dagger}$\\
            \hline
            \multirow{2}{*}{ASAP-Net} & 20.4 & 0.634 & 15.7 & 0.626 & 20.3 & 0.669 & 18.5 & 0.593 \\
            & $\pm 3.67^{\dagger}$ & $\pm 0.115^{\dagger}$ & $\pm 4.48$ & $\pm 0.103^{\dagger}$ & $\pm 3.05^{\dagger}$ & $\pm 0.089^{\dagger}$ & $\pm 3.60^{\dagger}$ & $\pm 0.086^{\dagger}$\\
            \hline
		CoNeS & $\bm{20.9}$ & $\bm{0.667}$ & 15.8 & $\bm{0.666}$ & $\bm{21.5}$ & $\bm{0.739}$ & $\bm{19.6}$ & $\bm{0.663}$ \\
            (proposed) & $\bm{\pm 3.66}$ & $\bm{\pm 0.099}$ & $\pm 4.44$ & $\pm 0.094$ & $\bm{\pm 3.35}$ & $\bm{\pm 0.095}$ & $\bm{\pm 3.49}$ & $\bm{\pm 0.084}$ 
		\end{tabular}
            \label{tb:bratsimagequality_cropped}
    \end{table}
    
     \paragraph{Benchmark overview} We compared our model with the following state-of-the-art methods: (1) pix2pix: pix2pix is a GAN-based image translation model, which consists of a UNet-based generator and a patch-based discriminator \citep{isola2017}; (2) pGAN: pGAN is a GAN-based image translation model using ResNet which follows an encoder-bottleneck-decoder architecture as backbone \citep{dar2019image}. Perceptual loss is introduced to improve the results; (3) ResViT: ResViT is an image translation model that combines pGAN with a transformer-based information bottleneck; (4) ASAP-Net: ASAP-Net is a neural field-based image translation model \citep{Shaham2021}. Different from the proposed model, ASAP-Net parameterizes patch-wise neural fields, which are conditioned through a UNet-shape hypernetwork without a shared MLP. For all implementations, we used the official GitHub repositories provided by the authors. We used the $\ell_1$ loss as a reconstruction loss for all the benchmark methods. We used the least square loss function \citep{mao2017least} as an adversarial loss for pix2pix, pGAN, and ResViT. Like the proposed method, we used the hinge loss function \citep{lim2017geometric} as an adversarial loss for ASAP-net. All the benchmark methods were trained using hyperparameters that were optimized by the original authors on the same dataset (BraTS). We trained ResViT with the pre-trained network as suggested in \citep{dalmaz2022resvit}, while all other models were trained from scratch.

    \begin{figure}[tb]
        \centering
        \begin{subfigure}{0.45\textwidth}
            \centering
            \caption{T1, T2, FLAIR $\rightarrow$ T1ce}
            \includegraphics[width=\textwidth]{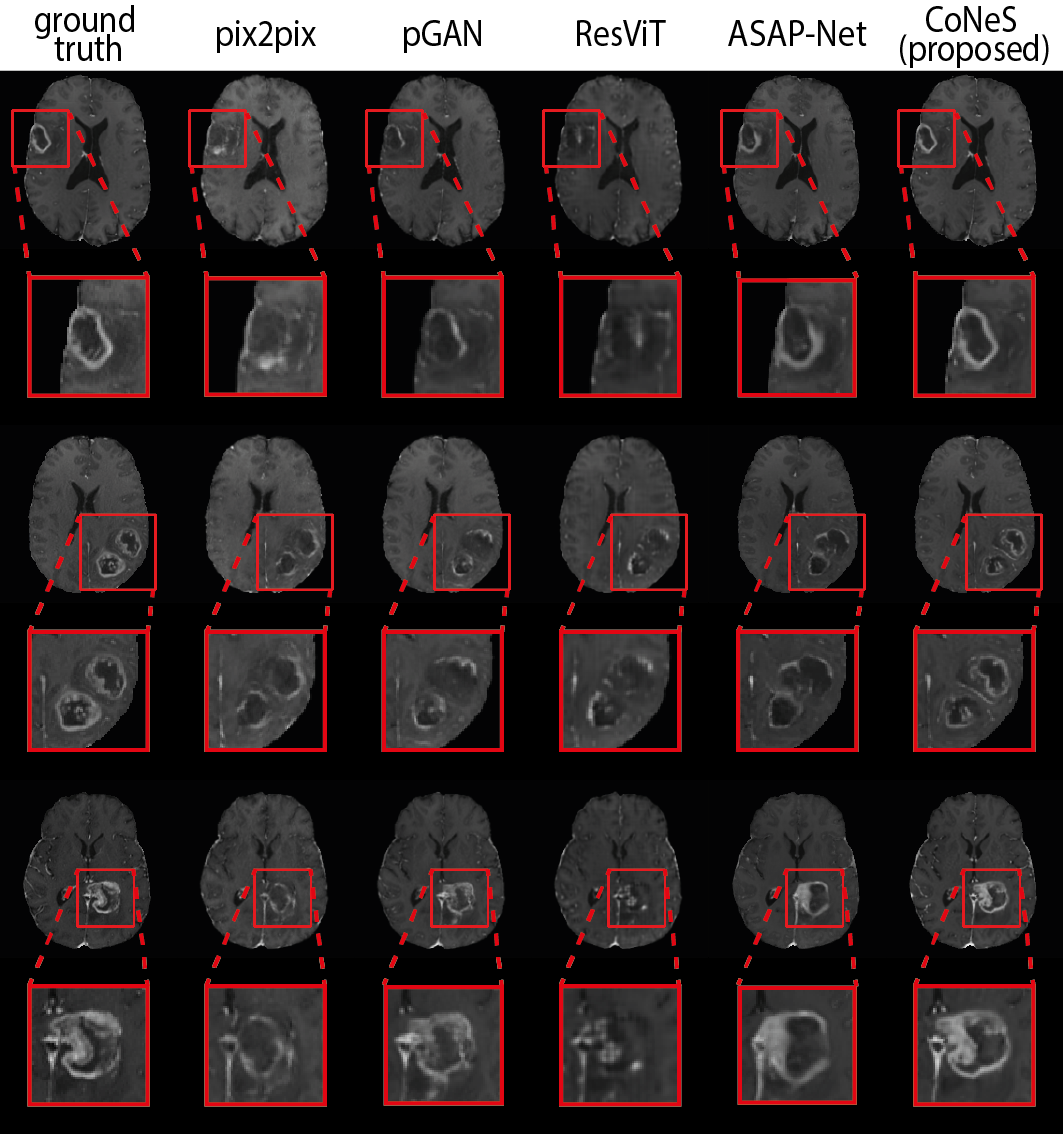}
            \label{fig:brats1}
            
        \end{subfigure}%
        \hspace{0.2em}
        \begin{subfigure}{0.45\textwidth}
            \centering
            \caption{T1ce, T2, FLAIR $\rightarrow$ T1}
            \includegraphics[width=\textwidth]{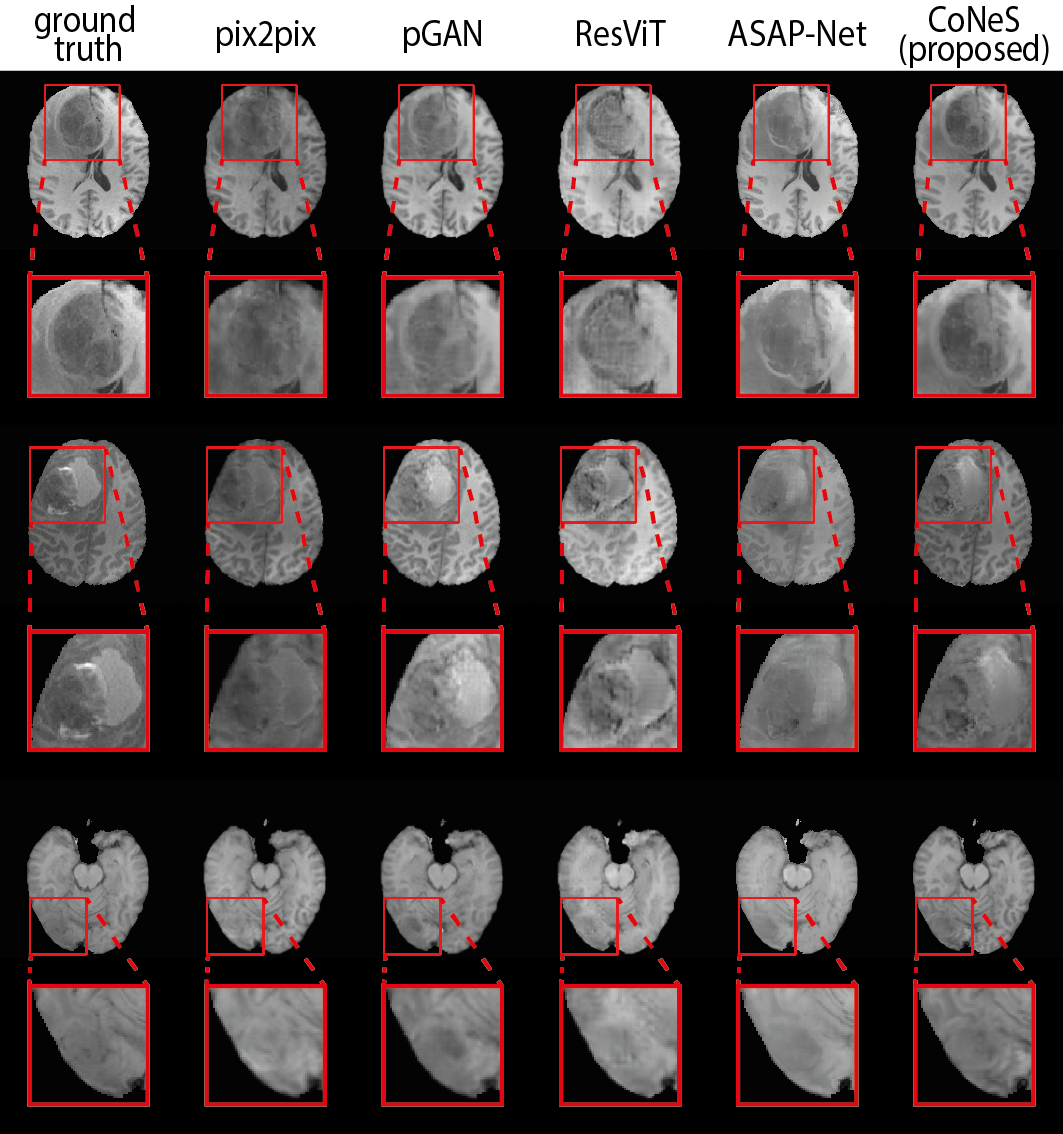}
            \label{fig:brats2}
        \end{subfigure}%
        \\[-3ex]
        \begin{subfigure}{0.45\textwidth}
            \centering
            \caption{T1ce, T1, FLAIR $\rightarrow$ T2}
            \includegraphics[width=\textwidth]{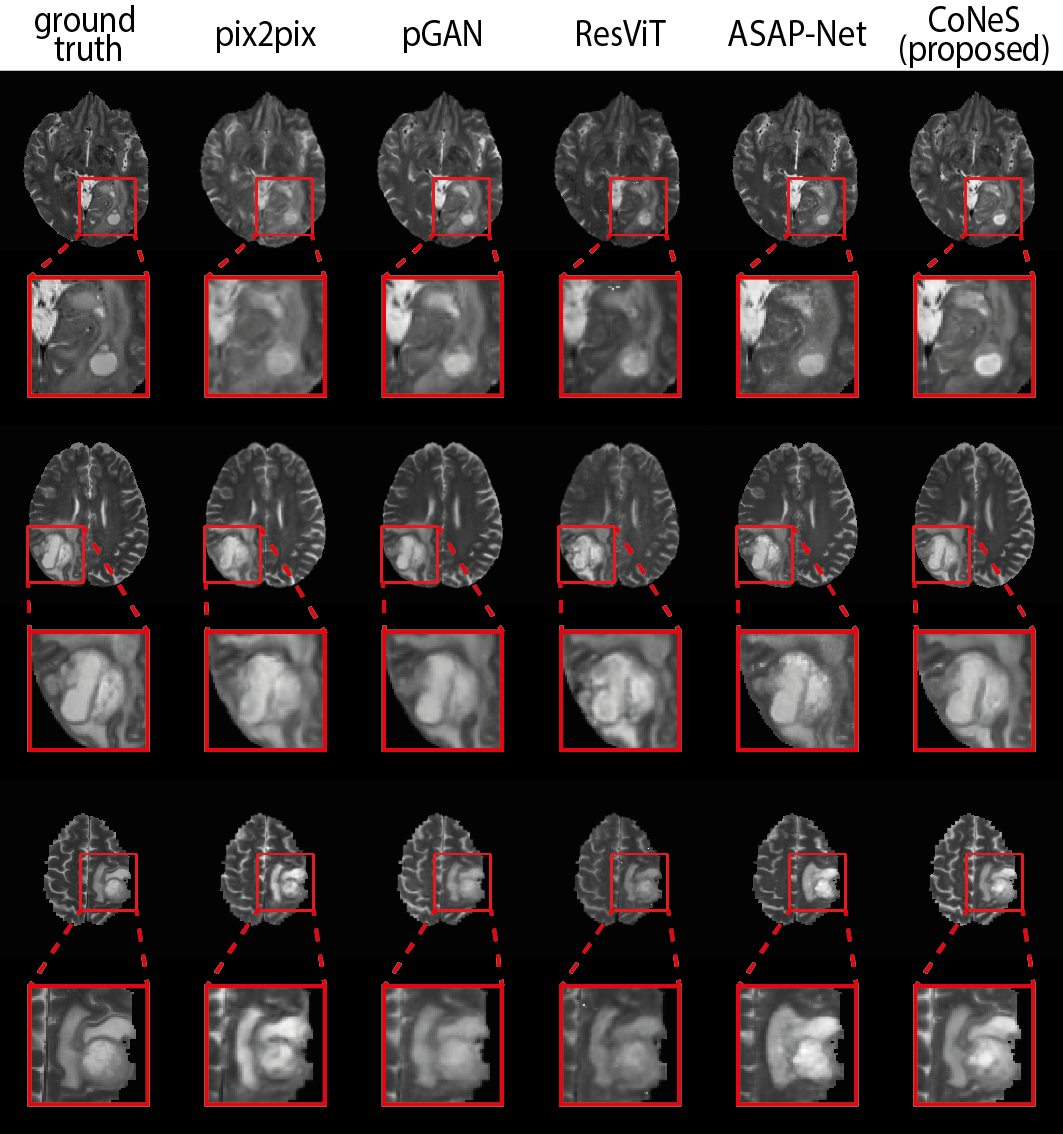}
            \label{fig:brats3}
        \end{subfigure}%
        \hspace{0.2em}
        \begin{subfigure}{0.45\textwidth}
        \centering
            \caption{T1ce, T1, T2 $\rightarrow$ FLAIR}
            \includegraphics[width=\textwidth]{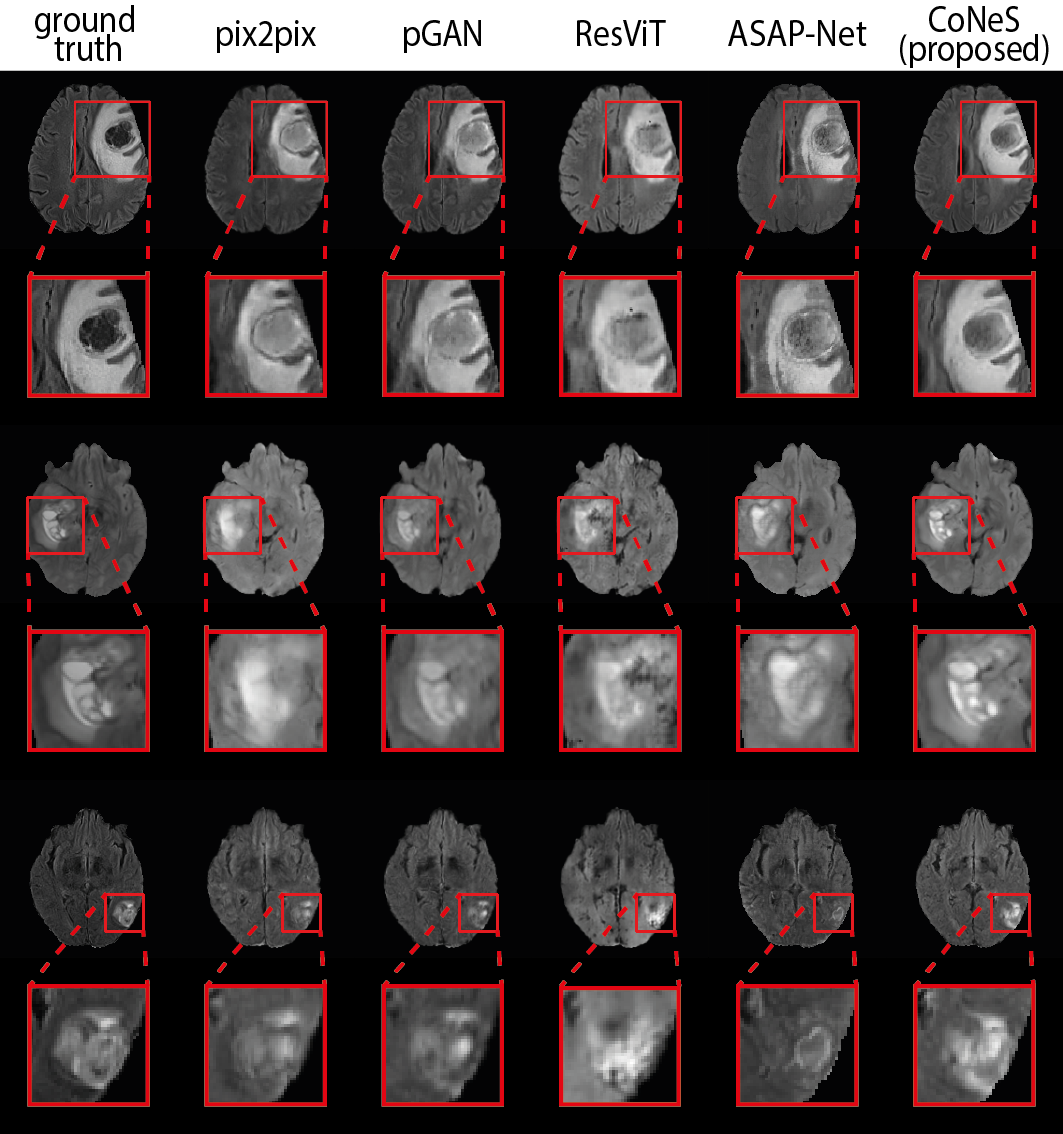}
            \label{fig:brats4}
        \end{subfigure}%
    \caption{Comparison results of different image translation models on BraTS 2018: (a) T1, T2, FLAIR $\rightarrow$ T1ce; (b) T1ce, T2, FLAIR $\rightarrow$ T1; (c) T1ce, T1, FLAIR $\rightarrow$ T2; (d) T1ce, T1, T2 $\rightarrow$ FLAIR. For each translation experiment, three examples are selected for display. Each column shows the ground truth and translation results of the different models. Zoomed-in results indicated in red rectangles are shown below the whole images.}
    \label{fig:bratssynthesis}
    \end{figure}
    
    \subsection{Multi-sequence MRI translation}
    We first examined the quality of the images generated from the proposed model. Theoretically, CoNeS can be applied to any number of missing or present sequences by adapting input and output channels to $N_s$ and $N_t$. For simplicity, we assumed one sequence was missing for all the patients during inference ($N_t = 1$), and thus, we trained models that generate one MRI sequence from the other sequences in the dataset for evaluation. Specifically, four image translation experiments were performed on BraTS 2018: (1) T1, T2, FLAIR $\rightarrow$ T1ce (shortened to T1ce translation); (2) T1ce, T2, FLAIR $\rightarrow$ T1 (shortened to T1 translation); (3) T1ce, T1, FLAIR $\rightarrow$ T2 (shortened to T2 translation); and (4) T1ce, T1, T2 $\rightarrow$ FLAIR (shortened to FLAIR translation). We used two different metrics for quantitative analysis in our study: peak signal-to-noise ratio (PSNR) and structural similarity index (SSIM). Both the synthesized images and real images were normalized to [0,1] before evaluation. Wilcoxon signed-rank test between each benchmark and the proposed model was performed on all image translation experiments.

    \begin{table}[tb] 
		\centering
		\caption{Quantitative comparison of different image translation models on VS dataset. The mean value and standard deviation of PSNR and SSIM are reported. The highest values per column are indicated in boldface; All metrics of the benchmarks in this table show significant differences (\(p < .05\)) compared to the proposed method.}
		\begin{tabular}{lcccccccc}
		\textbf{model} &  \multicolumn{2}{c}{\textbf{T1ce translation}} & \multicolumn{2}{c}{\textbf{T2 translation}}  \\
		\hline
            & PSNR & SSIM & PSNR & SSIM \\
            \hline
		pix2pix & $21.1 \pm 1.39$ & $0.602 \pm 0.068$ & $21.4 \pm 1.78$ & $0.506 \pm 0.121$ \\
            \hline
            pGAN & $21.6 \pm 1.55$ & $0.635 \pm 0.077$ & $22.2 \pm 2.04$ & $\bm{0.575 \pm 0.131}$\\
            \hline
            ResViT & $21.0 \pm 1.58$ & $0.575 \pm 0.090$ & $21.5 \pm 1.80$ & $0.489 \pm 0.110$\\
            \hline
            ASAP-Net & $20.4 \pm 1.24$ & $0.552 \pm 0.061$ & $20.9 \pm 1.97$ & $0.500 \pm 0.117$\\
            \hline
		CoNeS & \multirow{2}{*}{$\bm{21.9 \pm 1.69}$} & \multirow{2}{*}{$\bm{0.63 8 \pm 0.077}$} & \multirow{2}{*}{$\bm{22.6 \pm 2.03}$} & \multirow{2}{*}{$0.560 \pm 0.126$}  \\
            (proposed) &&&& \\
		\end{tabular}
            \label{tb:vsimagequality}
    \end{table}

    \begin{table}[tb] 
		\centering
		\caption{Quantitative comparison of different image translation models after cropping on VS dataset. The mean value and standard deviation of PSNR and SSIM are reported. The highest values per column are indicated in boldface; The $\dagger$ after each metric of the benchmarks indicates a significant difference (\(p < .05\)) compared to the proposed method.}
		\begin{tabular}{lcccccccc}
		\textbf{model} &  \multicolumn{2}{c}{\textbf{T1ce translation}} &  \multicolumn{2}{c}{\textbf{T2 translation}}\\
			\hline
                 & PSNR & SSIM & PSNR & SSIM & \\
                \hline
			pix2pix & $14.8 \pm 3.28$ & $0.415 \pm 0.122^{\dagger}$ & $16.6 \pm 1.72^{\dagger}$ & $0.321 \pm 0.084^{\dagger}$ \\
                \hline
                pGAN & $14.2 \pm 3.31^{\dagger}$ & $0.417 \pm 0.133^{\dagger}$ & $16.8 \pm 1.98^{\dagger}$ & $0.372 \pm 0.134$\\
                \hline
                ResViT & $14.5 \pm 2.75$ & $0.400 \pm 0.106^{\dagger}$ & $16.7 \pm 1.66^{\dagger}$ & $0.342 \pm 0.099^{\dagger}$ \\
                \hline
                ASAP-Net & $13.0 \pm 2.93^{\dagger}$ & $0.340 \pm 0.132^{\dagger}$ & $15.3 \pm 1.64^{\dagger}$ & $0.300 \pm 0.101^{\dagger}$\\
                \hline
			CoNeS & \multirow{2}{*}{$\bm{15.0 \pm 3.17}$} & \multirow{2}{*}{$\bm{0.451 \pm 0.118}$} & \multirow{2}{*}{$\bm{17.3 \pm 1.58}$} & \multirow{2}{*}{$\bm{0.379 \pm 0.101}$}\\ 
            (proposed) &&&& \\
		\end{tabular}
        \label{tb:vsimagequality_cropped}
    \end{table} 
    
    \begin{figure}[tb]
    \centering
    \begin{subfigure}{0.45\textwidth}
        \caption{T2 $\rightarrow$ T1ce}
        \includegraphics[width=\textwidth]{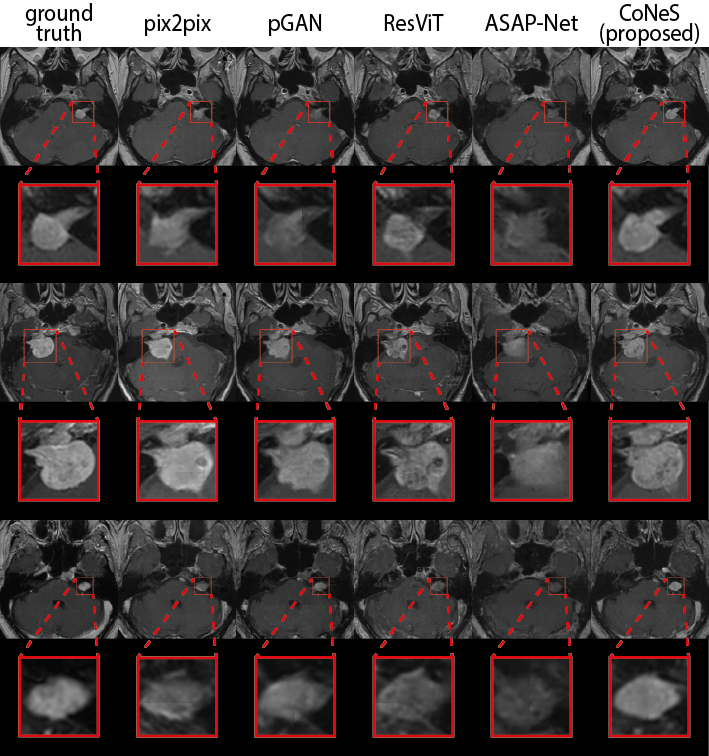}
    \end{subfigure}
    \hspace{0.2em}
    \begin{subfigure}{0.45\textwidth}
        \caption{T1ce $\rightarrow$ T2}
        \includegraphics[width=\textwidth]{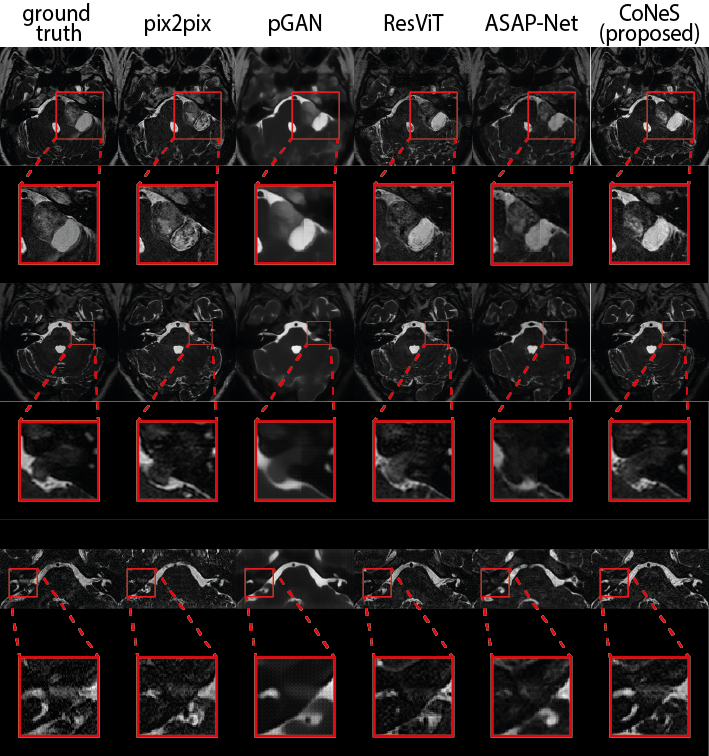}
    \end{subfigure}
    \caption{Comparison results of different image translation models on the VS dataset: (a) T2 $\rightarrow$ T1ce; (b) T1ce $\rightarrow$ T2. For each translation experiment, three examples are selected for display. Each column shows the ground truth and translation results of the different models. Zoomed-in results indicated in red rectangles are shown below the images.}
    \label{fig:vssynthesis}
    \end{figure}
    
    The quantitative results are listed in Table~\ref{tb:bratsimagequality}. As shown in the table, the proposed model performs significantly better ($p < .05$) than other state-of-the-art methods in most metrics, except that pGAN obtains higher PSNR on T1 translation. Using T1ce translation on BraTS 2018 as an example, the PSNR and SSIM of the proposed model on BraTS 2018 are 31.2 dB and 0.951, which outperforms pix2pix by 1.1 dB PSNR and 1.0\% SSIM, pGAN by 0.5 dB PSNR and 0.8\% SSIM, ResViT by 2.0 dB PSNR and 1.6\% SSIM, and ASAP-Net by 0.4 dB PSNR and 0.3\% SSIM. Translation examples are shown in Fig.~\ref{fig:bratssynthesis} in which we can see that the proposed model can recover more detailed structures, such as the contrast-enhanced tumor in T1ce, which is clinically highly relevant.
    
    Both the PSNR and SSIM show global similarity, while the quality of the region around the tumor is more clinically interesting. To further evaluate the proposed model, we cropped the images using the bounding box of the tumor region and then evaluated the similarity of these sub-images using the aforementioned metrics. The bounding box was generated from the segmentation results of nnUNet \citep{isensee2021} for the reason that the segmentation ground truths of BraTS 2018 validation set are not available. The results are listed in Table~\ref{tb:bratsimagequality_cropped}. As we can see, the proposed model also performs significantly better ($p<.05$) in most tasks within this sub-region, which is consistent with our observation from zoomed-in results in Fig.~\ref{fig:bratssynthesis}. We observed that the performance of the proposed model decreased after cropping due to the lack of background. Again using T1ce translation as an example, the PSNR and SSIM of the proposed model are 20.9 dB and 0.667, which outperforms pix2pix by 1.0 dB PSNR and 5.7\% SSIM, pGAN by 0.3 dB and 2.1\% SSIM, ResViT by 0.7 dB and 5.5\% SSIM, and ASAP-Net by 0.5 dB and 3.3\% SSIM. 
    
    Next, we performed two image translation experiments on the VS dataset: (1) T2 $\rightarrow$ T1ce (shortened to T1ce translation) and (2) T1ce $\rightarrow$ T2 (shortened to T2 translation). We again evaluated the entire image as well as the cropped region around the tumor, similar to BraTS 2018. Both quantitative results are listed in Table~\ref{tb:vsimagequality} and Table~\ref{tb:vsimagequality_cropped}. All models struggle with the VS dataset and show decreased performance compared to BraTS 2018, and CoNeS still performs significantly better ($p<.05$) in most of the metrics. Taking T1ce translation as an example, CoNeS obtains a PSNR of 21.9 dB and a SSIM score of 0.638, which outperforms pix2pix by 0.8 dB PSNR and 3.6\% SSIM, pGAN by 0.3 dB PSNR and 0.3\% SSIM, ResViT by 0.9 dB PSNR and 6.3\% SSIM, and ASAP-Net by 1.5 dB PSNR and 8.6\% SSIM. Qualitatively, we can observe improved synthesized images using the proposed model as shown in Fig.~\ref{fig:vssynthesis}. It is worth pointing out that although pGAN obtained better SSIM scores (0.575) in T2 translation, the visualization suggests that our results contain more informative details, while pGAN's results are blurry.  

    \begin{figure}[tb]
	\centering
    \begin{subfigure}[h]{\textwidth}
        \caption{Spectral analysis on BraTS 2018}
        \includegraphics[width=\columnwidth]{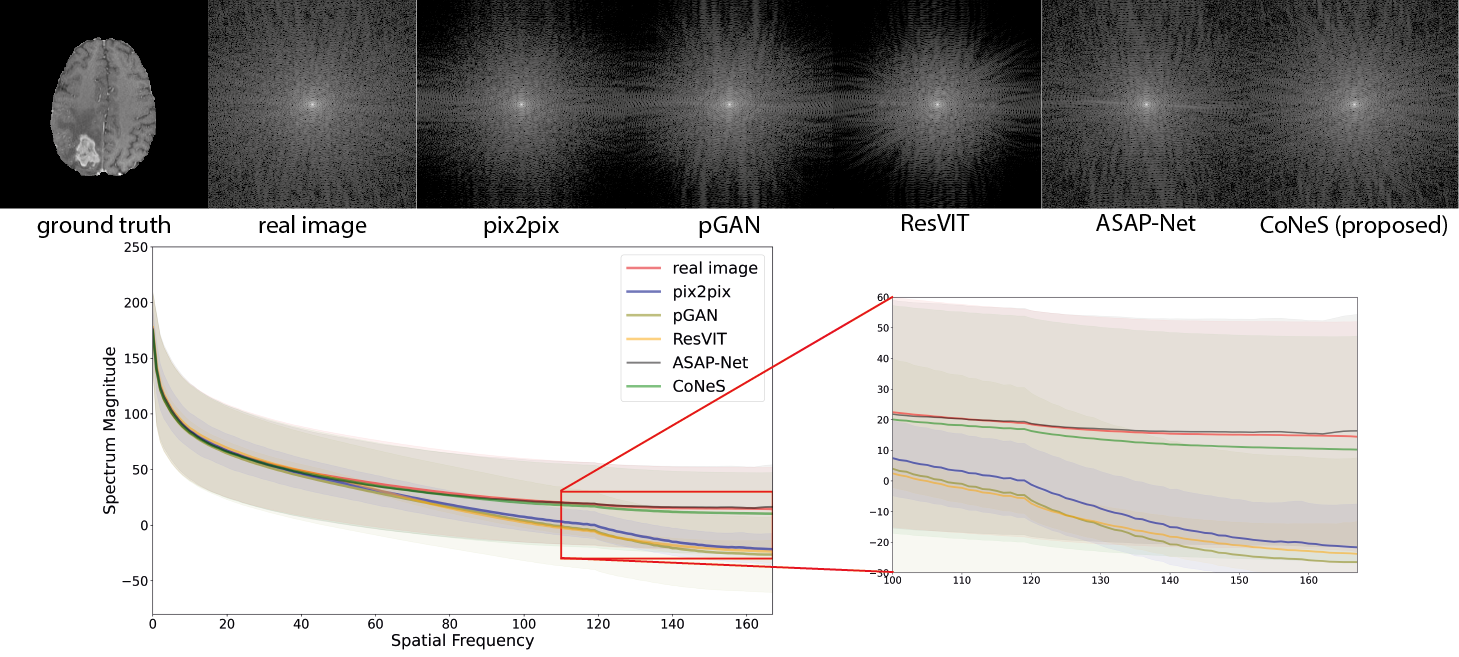}
        \label{fig:brats_spectrum}
    \end{subfigure}
    \\[-3ex]
    \begin{subfigure}[h]{\textwidth}
        \caption{Spectral analysis on the VS dataset}
	   \includegraphics[width=\columnwidth]{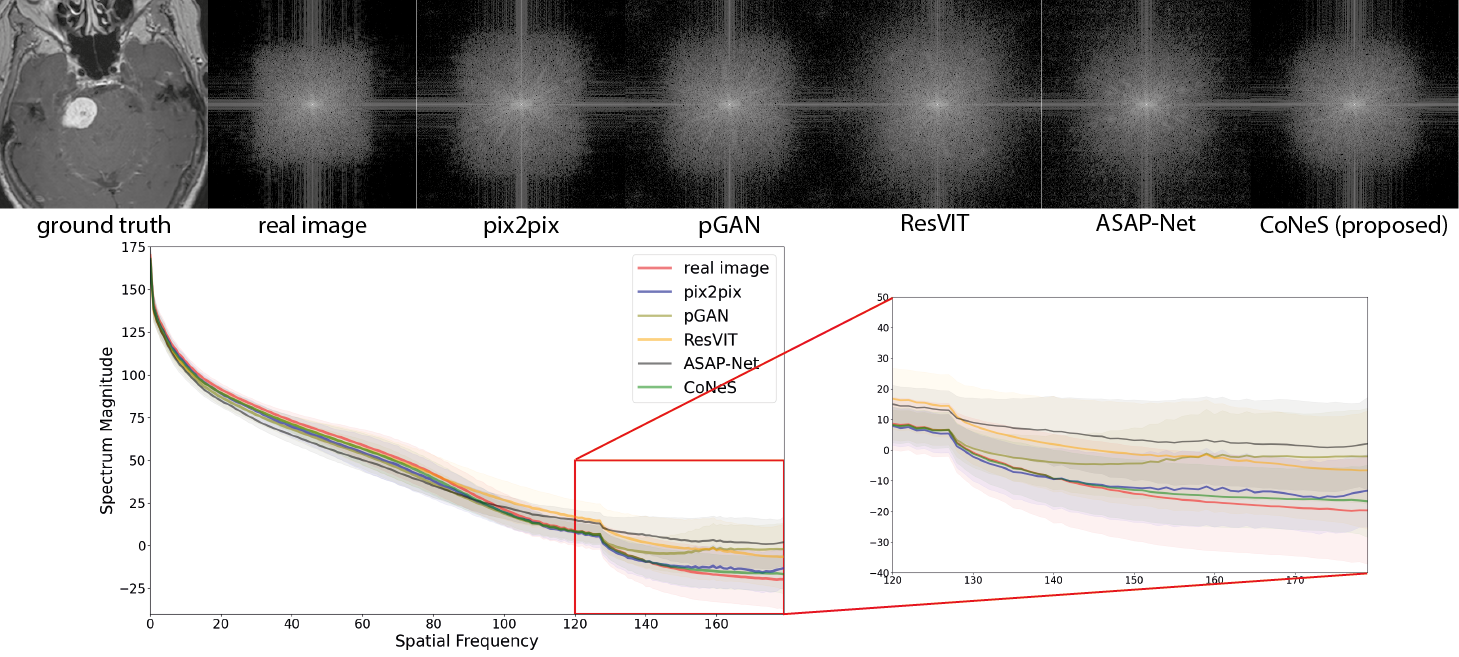}
        \label{fig:vs_spectrum}
    \end{subfigure}
	\caption{Spectral analysis of different image translation models. (a) and (b) show the analysis results on BraTS 2018 and the VS dataset, respectively. For each analysis, the Fourier transform of different synthesized images and the real image are shown in the top row. The bottom row shows the spectral distribution, in which the high-frequency range is zoomed in by the red rectangle.}
    \label{fig:spectral_analysis}
    \end{figure} 
    
    \subsection{Spectral analysis}
    Research has shown that CNN-based generative models with up-sampling layers usually struggle with reproducing the spectral distribution correctly \citep{durall2020watch,anokhin2021image}. On the contrary, coordinate-based networks like CoNeS build a direct pixel-to-pixel mapping without any up-sampling layer. In this section, we further evaluated the synthesized images in the frequency domain to demonstrate the improvement we obtained by performing spectral analysis on the T1ce translation model of both datasets. Specifically, we applied a 2D Fourier transform to all synthesized results as well as the real images, and then calculated a 1D representation of the 2D spectrum using Azimuthal integration \citep{durall2020watch}. Azimuthal integration is defined as an integration over the radial frequencies:
    \begin{linenomath}
    \begin{align}
        \text{AI}(\omega_k) &= \int_{0}^{2 \pi} \| \mathcal{F}(\omega_k \cos{\theta}, \omega_k \sin{\theta}) \omega_k\|_2 d\theta,
        \label{eq:ai}
    \end{align}
    \end{linenomath}
    for $k=0, \ldots, M/2-1$, and where $\mathcal{F}(m,n)$ is the Fourier Transform of a 2D image, $\theta$ is the radian, $\omega_k$ is the spatial frequency and $M$ is the length of a square image.
    
    A log transformation was performed to the 2D spectrum for better visualization, and we calculated the average 1D representation over the dataset to avoid biased sampling. As shown in Fig.~\ref{fig:spectral_analysis}, both ASAP-Net and CoNeS, which are coordinate-based networks, can reproduce the spectrum over the entire frequency range on BraTS 2018. Specifically, all the spectral curves are very close in the low-frequency range (spatial frequency $< 50$) which enables the generative models to reconstruct the general structure of the images. However, the spectral curves of GAN-based models dramatically drop in the high-frequency range (spatial frequency $> 75$), while the curves of ASAP-Net and CoNeS remain close to the real distribution. This shows that neural fields are able to overcome the spectral biases issue of convolutional neural networks. On the VS dataset, all the models yield higher spectrum magnitudes in the high-frequency range compared to the real images, which suggests that these translation models might add high-frequency noise to the synthesized images. Consistent with the similarity measurement results, ASAP-Net is not robust enough to reproduce the spectrum on the VS dataset and may induce more artifacts. On the contrary, CoNeS still outputs images whose spectrum is closest to the real images among all the translation models. The results indicate that by using neural fields conditioned via shift modulation, CoNeS is able to keep the representation capability and reproduce the spectrum distribution.
    
    \subsection{Synthesized images for tumor segmentation}
    To further examine the impact of synthesized images in downstream analysis, we performed tumor segmentation using the synthesized images at inference. To do this, we first adopted the architecture from nnUnet \citep{isensee2021} and trained a segmentation network that uses all the sequences in the dataset as input. Note that all the images were normalized to a range of [-1,1] during training to make the input channels consistent with the synthesized images. During inference, we tested the segmentation model with synthesized images and compared the results with the performance of the model when filling the missing channel with zeros, called zero imputation, in our experiments. For simplicity, we again assumed one specific sequence was missing and replaced this sequence while keeping the rest unchanged. Similar to the image translation experiments, we compared the segmentation performance using synthesized images generated from the proposed model to the other images via the Wilcoxon signed-rank test.
    
    \begin{table}[tb] 
	\centering
        \small
	\caption{Results of using different images for segmentation inference on BraTS 2018. The real sequences used are indicated by $\checkmark$, the missing ones by \ding{55}, and the ones replaced by synthesized images by \raisebox{0.8ex}{\protect\circled{}}. The mean of Dice scores and 95\% HD (mm) of the enhanced tumor (ET), the whole tumor (WT), and the tumor core (TC) are reported. The highest values per column are indicated in boldface; The $\dagger$ after each metric of the benchmarks indicates a significant difference (\(p < .05\)) compared to inference using synthesized images from CoNeS.}
        
	\begin{tabular}{cccc|c|c|c|c|c|c|c}
	\multicolumn{4}{c|}{input sequences} & \multirow{2}{*}{method} &  \multicolumn{3}{c|}{\textbf{Dice}} & \multicolumn{3}{c}{\textbf{95\% HD (mm)}}\\
	\cline{1-4}
        \cline{6-11}
        \multicolumn{1}{c|}{T1ce} & \multicolumn{1}{c|}{T1} & \multicolumn{1}{c|}{T2} & FLAIR && ET & WT  & TC  & ET  & WT  & TC  \\
        \hline
        \multicolumn{1}{c|}{\ding{51}} & \multicolumn{1}{c|}{\ding{51}} & \multicolumn{1}{c|}{\ding{51}}  & \ding{51} &  \diagbox{ }{ } & 0.770 & 0.888 & 0.822 & $4.49 $ & $6.27$  & $ 8.92$  \\
        \hline
        \multicolumn{1}{c|}{\ding{55}} & \multicolumn{1}{c|}{\ding{51}} & \multicolumn{1}{c|}{\ding{51}}  & \ding{51} &  zero imputation & $0.068^{\dagger}$ & $0.845^{\dagger}$ & $0.362^{\dagger}$ & $27.9^{\dagger}$ & $8.80^{\dagger}$  & $18.1^{\dagger}$  \\
        \multicolumn{1}{c|}{\raisebox{\depth}{\circled}}    & \multicolumn{1}{c|}{\ding{51}}   & \multicolumn{1}{c|}{\ding{51}}   & \ding{51} &  pix2pix & $0.191^{\dagger}$ & $0.850^{\dagger}$ & $0.537^{\dagger}$ & $15.0^{\dagger}$ & $8.10^{\dagger}$  & $15.0^{\dagger}$ \\
        \multicolumn{1}{c|}{\raisebox{\depth}{\circled}}    & \multicolumn{1}{c|}{\ding{51}}   & \multicolumn{1}{c|}{\ding{51}}   & \ding{51} &  pGAN & $0.317^{\dagger}$ & $0.858^{\dagger}$ & $0.598^{\dagger}$ & $14.9^{\dagger}$ & $8.01^{\dagger}$ & $14.0^{\dagger}$\\
        \multicolumn{1}{c|}{\raisebox{\depth}{\circled}}    & \multicolumn{1}{c|}{\ding{51}}   & \multicolumn{1}{c|}{\ding{51}}   & \ding{51} &  ResViT & $0.223^{\dagger}$ & $0.858^{\dagger}$ & $0.555^{\dagger}$ &  $15.0^{\dagger}$ & $7.87^{\dagger}$ & $14.1^{\dagger}$ \\
        \multicolumn{1}{c|}{\raisebox{\depth}{\circled}}    & \multicolumn{1}{c|}{\ding{51}}   & \multicolumn{1}{c|}{\ding{51}}   & \ding{51} &  ASAP-Net & $0.332^{\dagger}$& $0.866^{\dagger}$ & $0.597^{\dagger}$ & $13.3^{\dagger}$ & $\bm{6.95}$ & $13.2^{\dagger}$\\
	\multicolumn{1}{c|}{\raisebox{\depth}{\circled}}    & \multicolumn{1}{c|}{\ding{51}}   & \multicolumn{1}{c|}{\ding{51}}   & \ding{51} & CoNeS & $\bm{0.386}$ & $\bm{0.870}$ & $\bm{0.662}$ & $\bm{13.1}$ & $7.23$ & $\bm{13.0}$ \\
	\hline
        \multicolumn{1}{c|}{\ding{51}} & \multicolumn{1}{c|}{\ding{55}} & \multicolumn{1}{c|}{\ding{51}}  & \ding{51}  &  zero imputation & $0.717^{\dagger}$ & $0.865^{\dagger}$ & $0.753^{\dagger}$ & $6.53^{\dagger}$ & $7.86^{\dagger}$ & $11.3^{\dagger}$ \\
        \multicolumn{1}{c|}{\ding{51}} & \multicolumn{1}{c|}{\raisebox{\depth}{\circled}} & \multicolumn{1}{c|}{\ding{51}}  & \ding{51}  &  pix2pix & $0.747$ & $0.869^{\dagger}$ & $0.780^{\dagger}$ & $5.07^{\dagger}$ & $7.70^{\dagger}$ & $9.84^{\dagger}$ \\
        \multicolumn{1}{c|}{\ding{51}} & \multicolumn{1}{c|}{\raisebox{\depth}{\circled}} & \multicolumn{1}{c|}{\ding{51}}  & \ding{51}  &  pGAN & $0.747^{\dagger}$ & $0.868^{\dagger}$ & $0.779^{\dagger}$ & $5.60^{\dagger}$ & $7.61^{\dagger}$ & $10.2^{\dagger}$ \\
        \multicolumn{1}{c|}{\ding{51}} & \multicolumn{1}{c|}{\raisebox{\depth}{\circled}} & \multicolumn{1}{c|}{\ding{51}}  & \ding{51}  &  ResViT & $0.751^{\dagger}$ & $0.869^{\dagger}$ & $0.784^{\dagger}$ & $\bm{4.62}$ & $7.39^{\dagger}$ & $9.75^{\dagger}$ \\
        \multicolumn{1}{c|}{\ding{51}} & \multicolumn{1}{c|}{\raisebox{\depth}{\circled}} & \multicolumn{1}{c|}{\ding{51}}  & \ding{51}  &  ASAP-Net & $0.753^{\dagger}$ & 0.881 & $0.806$ & $5.43^{\dagger}$ & $\bm{6.73}$ & 9.39\\
        \multicolumn{1}{c|}{\ding{51}} & \multicolumn{1}{c|}{\raisebox{\depth}{\circled}} & \multicolumn{1}{c|}{\ding{51}}  & \ding{51}  &  CoNeS & $\bm{0.764}$ & $\bm{0.885}$ & $\bm{0.808}$ & $5.30$ & $7.05$ & $\bm{8.94}$ \\
        \hline
        \multicolumn{1}{c|}{\ding{51}} & \multicolumn{1}{c|}{\ding{51}} & \multicolumn{1}{c|}{\ding{55}}  & \ding{51}  &  zero imputation & $0.748^{\dagger}$ & $0.835^{\dagger}$ & $0.752^{\dagger}$ & $5.64^{\dagger}$ & $8.67^{\dagger}$ & $11.6^{\dagger}$ \\
        \multicolumn{1}{c|}{\ding{51}} & \multicolumn{1}{c|}{\ding{51}} & \multicolumn{1}{c|}{\raisebox{\depth}{\circled}}  & \ding{51}  &  pix2pix & $0.761$ & $0.862^{\dagger}$ & $0.784^{\dagger}$ & $3.90^{\dagger}$ & $7.53^{\dagger}$ & $9.82^{\dagger}$ \\
        \multicolumn{1}{c|}{\ding{51}} & \multicolumn{1}{c|}{\ding{51}} & \multicolumn{1}{c|}{\raisebox{\depth}{\circled}}  & \ding{51}  &  pGAN & $0.767$ & $0.872^{\dagger}$ & $0.797^{\dagger}$ & $3.83^{\dagger}$ & $7.34^{\dagger}$ & $9.27$ \\
        \multicolumn{1}{c|}{\ding{51}} & \multicolumn{1}{c|}{\ding{51}} & \multicolumn{1}{c|}{\raisebox{\depth}{\circled}}  & \ding{51}  &  ResViT & $0.759$ & $0.855^{\dagger}$ & $0.788^{\dagger}$ & $4.19^{\dagger}$ & $8.18^{\dagger}$ & $9.74 $ \\
        \multicolumn{1}{c|}{\ding{51}} & \multicolumn{1}{c|}{\ding{51}} & \multicolumn{1}{c|}{\raisebox{\depth}{\circled}}  & \ding{51}  &  ASAP-Net & 0.764 & $0.880^{\dagger}$ & 0.817 & $3.84^{\dagger}$ & $6.50^{\dagger}$ & 9.05\\
        \multicolumn{1}{c|}{\ding{51}} & \multicolumn{1}{c|}{\ding{51}} & \multicolumn{1}{c|}{\raisebox{\depth}{\circled}}  & \ding{51}  &  CoNeS & $\bm{0.778}$ & $\bm{0.886}$ & $\bm{0.829}$ & $\bm{3.15}$ & $\bm{6.01}$ & $\bm{8.34}$ \\
        \hline
        \multicolumn{1}{c|}{\ding{51}} & \multicolumn{1}{c|}{\ding{51}} & \multicolumn{1}{c|}{\ding{51}}  & \ding{55}  &  zero imputation & $0.679^{\dagger}$ & $0.403^{\dagger}$ & $0.690^{\dagger}$ &  $27.8^{\dagger}$ &  $30.3^{\dagger}$ & $23.2^{\dagger}$ \\
        \multicolumn{1}{c|}{\ding{51}} & \multicolumn{1}{c|}{\ding{51}} & \multicolumn{1}{c|}{\ding{51}}  & \raisebox{\depth}{\circled}  &  pix2pix & $0.760$ & $0.805^{\dagger}$ & $0.771^{\dagger}$ & $\bm{3.74}$ & $9.75^{\dagger}$ & $11.1^{\dagger}$ \\
        \multicolumn{1}{c|}{\ding{51}} & \multicolumn{1}{c|}{\ding{51}} & \multicolumn{1}{c|}{\ding{51}}  & \raisebox{\depth}{\circled}  &  pGAN & $0.766$ & $0.833^{\dagger}$ & $0.777^{\dagger}$ & $4.60$ & $8.59^{\dagger}$ & $10.3^{\dagger}$ \\
        \multicolumn{1}{c|}{\ding{51}} & \multicolumn{1}{c|}{\ding{51}} & \multicolumn{1}{c|}{\ding{51}}  & \raisebox{\depth}{\circled}  &  ResViT & $0.783$ & $0.768^{\dagger}$ & $0.752^{\dagger}$ & $5.16^{\dagger}$ & $11.7^{\dagger}$ & $11.5^{\dagger}$ \\
        \multicolumn{1}{c|}{\ding{51}} & \multicolumn{1}{c|}{\ding{51}} & \multicolumn{1}{c|}{\ding{51}}  & \raisebox{\depth}{\circled}  &  ASAP-Net & $\bm{0.785}$ & $0.823^{\dagger}$ & $0.808^{\dagger}$ & $3.80^{\dagger}$ & $9.36^{\dagger}$ & $\bm{9.36}^{\dagger}$\\
        \multicolumn{1}{c|}{\ding{51}} & \multicolumn{1}{c|}{\ding{51}} & \multicolumn{1}{c|}{\ding{51}}  & \raisebox{\depth}{\circled}  &  CoNeS & $0.768$ & $\bm{0.853}$ & $\bm{0.809}$ & $4.30$ & $\bm{7.56}$ & $9.38 $ \\
        \end{tabular}
        \label{tb:bratsseg}
    \end{table}

    \begin{figure}[tb]
        \centering
        \begin{subfigure}[h]{\textwidth}
            \caption{Segmentation example on BraTS 2018}
            \includegraphics[width=\textwidth]{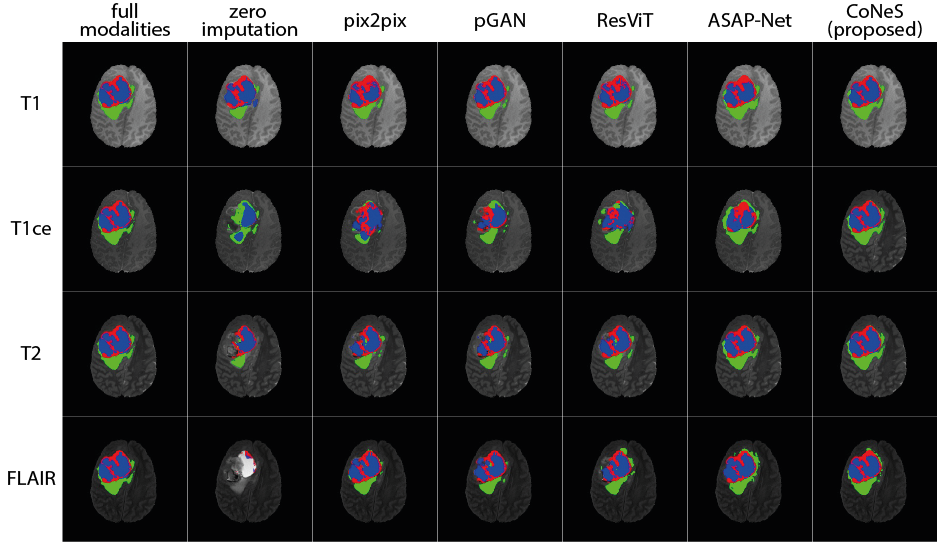} 
        \end{subfigure}
        \begin{subfigure}[h]{\textwidth}
            \caption{Segmentation example on the VS dataset}
            \includegraphics[width=\textwidth]{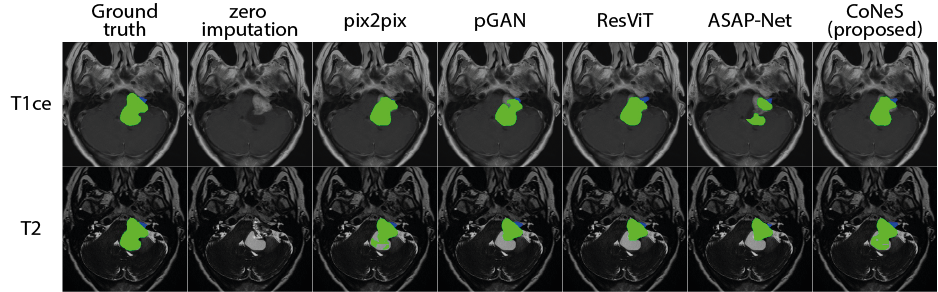}
        \end{subfigure}
    \caption{The results of segmentation experiments: (a) A segmentation example on BraTS 2018 and (b) an example on the VS dataset. The rows show the segmentation results with different MRI sequences replaced. The columns show ground truth (for BraTS 2018, segmentation results with full sequences) and segmentation results using different synthesized images.}
    \label{fig:brats_seg}
    \end{figure}
    
    The tests were performed for each MRI sequence (T1ce, T1, T2, and FLAIR) on BraTS 2018. The performance was evaluated using three specific categories: 1) enhanced tumor (ET); 2) tumor core (TC, non-enhanced tumor, and edema); and 3) the whole tumor (WT, enhanced tumor, non-enhanced tumor, and edema). Dice score and 95\% Hausdorff distance (95\% HD) of all the three categories are reported for quantitative evaluation in Table~\ref{tb:bratsseg}. We can see that the presence of sequences dramatically influences the performance of the segmentation model. For instance, when the T1ce is missing, the Dice score of the enhanced tumor is 0.068 because the enhanced information is only visible in the T1ce. As expected, most of the metrics show that inference with synthesized images performs worse than inference with full sequences. However, we also noticed that when the real T2 or FLAIR were replaced with synthesized ones, we obtained a lower mean 95\% HD. This occurs due to the influence of certain outliers. For example, sometimes the model can identify the enhanced tumor at the wrong position using real images, leading to a large 95\% HD, while the other inferences using synthesized images completely miss the tumor. When we removed three outliers, the mean of 95\% HD of the enhanced tumor became 2.99 mm, which is better than the others. 
    
    The best results among all the inferences using synthesized images (including zero imputation) for each sequence were highlighted in Table~\ref{tb:bratsseg}. The results indicate that using synthesized images for inference can significantly improve the segmentation performance and that the synthesized images of our model yield the best segmentation performance with a significant difference ($p < .05$) among all the translation models. Using the inferences without T1ce as examples, the Dice scores of the proposed model are 0.386, 0.870, and 0.662 in the enhanced tumor, the whole tumor, and the tumor core respectively. In comparison, the proposed model outperforms zero imputation by 31.8\%, 2.5\%, and 30.0\%, pix2pix by 19.5\%, 2.0\%, and 12.5\%, pGAN by 6.9\%, 1.2\%, and 6.4\%, ResViT by 16.3\%, 1.2\%, and 10.7\%, and ASAP-Net by 5.4\%, 0.4\%, and 6.5\%. The 95\% HDs of the proposed model are 13.1 mm, 7.23 mm, and 13.0 mm in the enhanced tumor, the whole tumor, and the tumor core respectively. In comparison, the proposed model outperforms zero imputation by 14.8 mm, 1.57 mm, and 5.1 mm, pix2pix by 1.9 mm, 0.87 mm, and 2.0 mm, pGAN by 1.8 mm, 0.78 mm, 1.0 mm, ResViT by 1.9 mm, 0.64 mm, 1.1 mm. Although ASAP-Net obtained a higher 95\% HD (6.95 mm) in the whole tumor, we did not observe significant differences between it and the proposed model. Some example segmentation results are presented in Fig.~\ref{fig:brats_seg}. It is worth noting that the synthesized T1 of CoNeS performs better in segmentation than the ones from pGAN, although we got higher PSNR for pGAN in the former experiment.

    We also performed the same segmentation experiments on the VS dataset. We evaluated the performance using three specific categories: 1) intrameatal tumor; 2) extrameatal tumor); and 3) the whole tumor (including intra- and extrameatal tumor). Dice score and 95\% HD of all three categories are reported in Table~\ref{tb:vsseg}. Similarly to BraTS 2018, all the synthesized images compensate for the performance loss due to the drop of sequences, and the proposed model performs significantly better ($p < .05$) than the other models. For instance, the synthesized T1ce generated by the proposed model obtained Dice scores of 0.567, 0.714, and 0.749 in the intrameatal tumor, the extrameatal tumor, and the whole tumor respectively. In comparison, the proposed model outperforms zero imputation by 56.6\%, 68.4\%, and 72.1\%, pix2pix by 9.6\%, 2.8\%, and 3.6\%, pGAN by 12.6\%, 3.8\%, and 8.8\%, ResViT by 2.7\%, 0.1\%, and 3.0\%, and ASAP-Net by 25.9\%, 22.2\%, and 24.1\%. The 95\% HDs of the proposed model are 2.33 mm, 3.54 mm, and 4.05 mm in the intrameatal tumor, the extrameatal tumor, and the whole tumor respectively. These results outperform zero imputation by 5.71 mm, 25.37 mm, and 30.05, pix2pix by 0.21 mm, 2.13 mm, and 2.45 mm, pGAN by 0.15 mm, 3.03 mm, and 3.82 mm, ASAP-Net by 0.77 mm, 3.72 mm, and 8.35 mm. We observed that ResViT obtained lower 95\% HD (3.32 mm) in the extrameatal tumor, however, the proposed model still performs better than ResViT in most of the experiments. Example segmentation results are displayed in Fig.~\ref{fig:brats_seg}. 

    \begin{table}[tb] 
        \centering
	  \caption{Results of using different images for segmentation inference on the VS dataset. The real sequences used are indicated by $\checkmark$, the missing ones by \ding{55}, and the ones replaced by synthesized images by \raisebox{0.8ex}{\protect\circled{}}. The mean of Dice scores and 95\% HD (mm) of the intrameatal tumor (IT), the extrameatal tumor (ET), and the whole tumor (WT) are reported. The highest values per column are indicated in boldface; The $\dagger$ after each metric of the benchmarks indicates significant differences (\(p < .05\)) compared to inference using synthesized images from CoNeS.}
        \begin{tabular}{cc|c|c|c|c|c|c|c}
	   \multicolumn{2}{c|}{input sequences} & \multirow{2}{*}{method} &  \multicolumn{3}{c|}{\textbf{Dice}} & \multicolumn{3}{c}{\textbf{95\% HD (mm)}}\\
       \cline{1-2}
       \cline{4-9}
        \multicolumn{1}{c|}{T1ce} & \multicolumn{1}{c|}{T2} & & IT & ET  & WT  & IT  & ET  & WT \\
        \hline
        \multicolumn{1}{c|}{\ding{51}} & \multicolumn{1}{c|}{\ding{51}} & \diagbox{ }{ } & 0.761 & 0.853 & 0.896 & 1.34 & 1.71 & 1.45 \\
        \hline
        \multicolumn{1}{c|}{\ding{55}} & \multicolumn{1}{c|}{\ding{51}} & zero imputation & $0.001^{\dagger}$ & $0.030^{\dagger}$ & $0.028^{\dagger}$ & $8.04^{\dagger}$ & $28.91^{\dagger}$ & $34.1^{\dagger}$ \\
         \multicolumn{1}{c|}{\raisebox{\depth}{\circled}}  & \ding{51} & pix2pix & $0.471^{\dagger}$ & $0.686^{\dagger}$ & $0.713^{\dagger}$ & 2.54 & $5.67^{\dagger}$ & 6.50\\
         \multicolumn{1}{c|}{\raisebox{\depth}{\circled}}  & \ding{51} & pGAN & $0.441^{\dagger}$ & $0.676^{\dagger}$ & $0.661^{\dagger}$ & $2.48^{\dagger}$ & $6.57^{\dagger}$ & $7.87^{\dagger}$ \\
         \multicolumn{1}{c|}{\raisebox{\depth}{\circled}}  & \ding{51} & ResViT & $0.540^{\dagger}$ & $0.713^{\dagger}$ & $0.719$ & 2.36 & $\bm{3.32}^{\dagger}$ & 5.73\\
         \multicolumn{1}{c|}{\raisebox{\depth}{\circled}}  & \ding{51} & ASAP-Net &  $0.308^{\dagger}$ & $0.492^{\dagger}$ & $0.508^{\dagger}$ & $3.10^{\dagger}$ & $7.26^{\dagger}$ & $12.4^{\dagger}$ \\
         \multicolumn{1}{c|}{\raisebox{\depth}{\circled}}  & \ding{51} & CoNeS & $\bm{0.567}$ & $\bm{0.714}$ & $\bm{0.749}$ & $\bm{2.33}$ & 3.54 & $\bm{4.05}$\\
         \hline
         \multicolumn{1}{c|}{\ding{51}} & \multicolumn{1}{c|}{\ding{55}} & zero imputation & $0.184^{\dagger}$ & $0.397^{\dagger}$ & $0.400^{\dagger}$ & $4.06^{\dagger}$ & $18.0^{\dagger}$ & $22.2^{\dagger}$\\
         \multicolumn{1}{c|}{\ding{51}} & \multicolumn{1}{c|}{\raisebox{\depth}{\circled}} & pix2pix & $0.713^{\dagger}$ & $0.856^{\dagger}$ & $0.874^{\dagger}$ & 1.54 & $2.19^{\dagger}$ & 2.09\\
         \multicolumn{1}{c|}{\ding{51}} & \multicolumn{1}{c|}{\raisebox{\depth}{\circled}} & pGAN & $0.701^{\dagger}$ & $0.839^{\dagger}$ & $0.844^{\dagger}$ & $1.82^{\dagger}$ & $2.60^{\dagger}$ & $2.58^{\dagger}$\\
         \multicolumn{1}{c|}{\ding{51}} & \multicolumn{1}{c|}{\raisebox{\depth}{\circled}} & ResViT & $0.716^{\dagger}$ & $0.831^{\dagger}$ & 0.862 & 1.61 & $2.48^{\dagger}$ & 2.32\\
         \multicolumn{1}{c|}{\ding{51}} & \multicolumn{1}{c|}{\raisebox{\depth}{\circled}} & ASAP-Net & $0.677^{\dagger}$ & $0.834^{\dagger}$ & $0.854^{\dagger}$ & $1.95^{\dagger}$ & $2.63^{\dagger}$ & $2.45^{\dagger}$ \\
         \multicolumn{1}{c|}{\ding{51}} & \multicolumn{1}{c|}{\raisebox{\depth}{\circled}} & CoNeS & $\bm{0.746}$ & $\bm{0.858}$ & $\bm{0.878}$ & $\bm{1.40}$ & $\bm{2.09}$ & $\bm{1.96}$ \\
        \end{tabular}
        \label{tb:vsseg}
    \end{table}

    \begin{figure}[tb]
        \centering
        \includegraphics[width=0.8\textwidth]{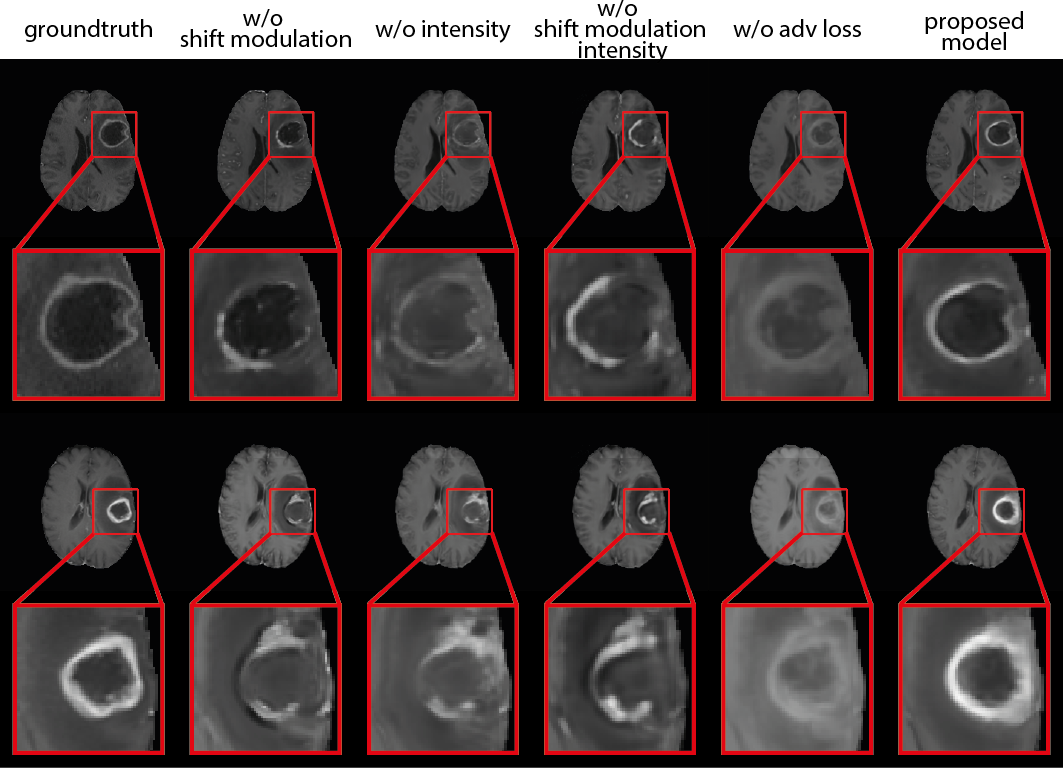} 
    \caption{Example results of the ablated models. Zoomed-in results indicated with red rectangles are shown below the full images.}
    \label{fig:ablation experiments}
    \end{figure}

    \subsection{Ablation study}
    Ablation studies were performed to verify the benefits of individual components in the proposed method. For simplicity, we trained a baseline CoNeS model that translates T1ce from T2 on BraTS 2018. We first examined the added value of the source image as input to the MLP by removing the intensity value from the input channel. In this case, the neural fields are conditioned on the latent code only. Next, we compared shift modulation against a full hypernetwork where all the parameters of the MLP are generated. Last, we trained the proposed method without the adversarial loss to show the contribution of the discriminator in our model. Wilcoxon signed-rank tests were performed between the baseline model and ablated models. Quantitative and qualitative results are shown in Table~{\ref{tb:ablation study}} and Fig.~{\ref{fig:ablation experiments}}. We noticed that although the model without adversarial loss achieves marginally better SSIM and PSNR, the results, especially the tumor region, are visually blurry, which shows that the adversarial loss helps the model to reconstruct more details and outputs more realistic images. Apart from this, the proposed model obtained the best results among the ablated models that include the adversarial loss and showed significant differences ($p < .01$) in SSIM. These results also show that shift modulation helps to reduce the parameters from 14.5k to 0.26k, which is the number of neurons in the MLP, without loss of representation capability. Moreover, although the instance-specific information is already encoded in the latent code, conditioning the network on intensity directly can still add extra information and improve performance.

    \begin{table}[tb] 
	\centering
	\caption{Quantitative comparison of ablated models on BraTS 2018. The mean value and standard deviation of PSNR and SSIM are reported. The highest values per column are indicated in boldface; The $\dagger$ after each metric indicates a significant difference ($p < .01$) compared to the proposed model (the bottom row).}
	\begin{tabular}{cccccc}
		shift & \multirow{2}{*}{intensity} & adversarial  & \#param & \multirow{2}{*}{PSNR} & \multirow{2}{*}{SSIM} \\
            modulation & &loss & generated& & \\
            \hline
             no & no & yes & 14.5k & $29.6 \pm 2.13^{\dagger}$ & $0.933 \pm 0.013^{\dagger}$ \\
            \hline
		   no & yes & yes & 14.5k & $29.9 \pm 2.32$ & $0.938 \pm 0.013^{\dagger}$\\
            \hline
		   yes & no & yes & 0.26k & $30.0 \pm 2.13$ & $0.938 \pm 0.013^{\dagger}$\\
            \hline
            yes & yes & no & 0.26k & $\bm{30.2 \pm 2.33}$ & $\bm{0.943  \pm 0.013}^{\dagger}$ \\
            \hline\hline
		  yes & yes & yes & 0.26k & $30.0 \pm 2.22$ & $0.941 \pm 0.014$
            \\
	\end{tabular}
        \label{tb:ablation study}
    \end{table}
    
\begin{figure}[tb]
        \centering
        \begin{subfigure}[h]{0.49\textwidth}
            \caption{Adversarial loss curves}
            \includegraphics[width=\textwidth]{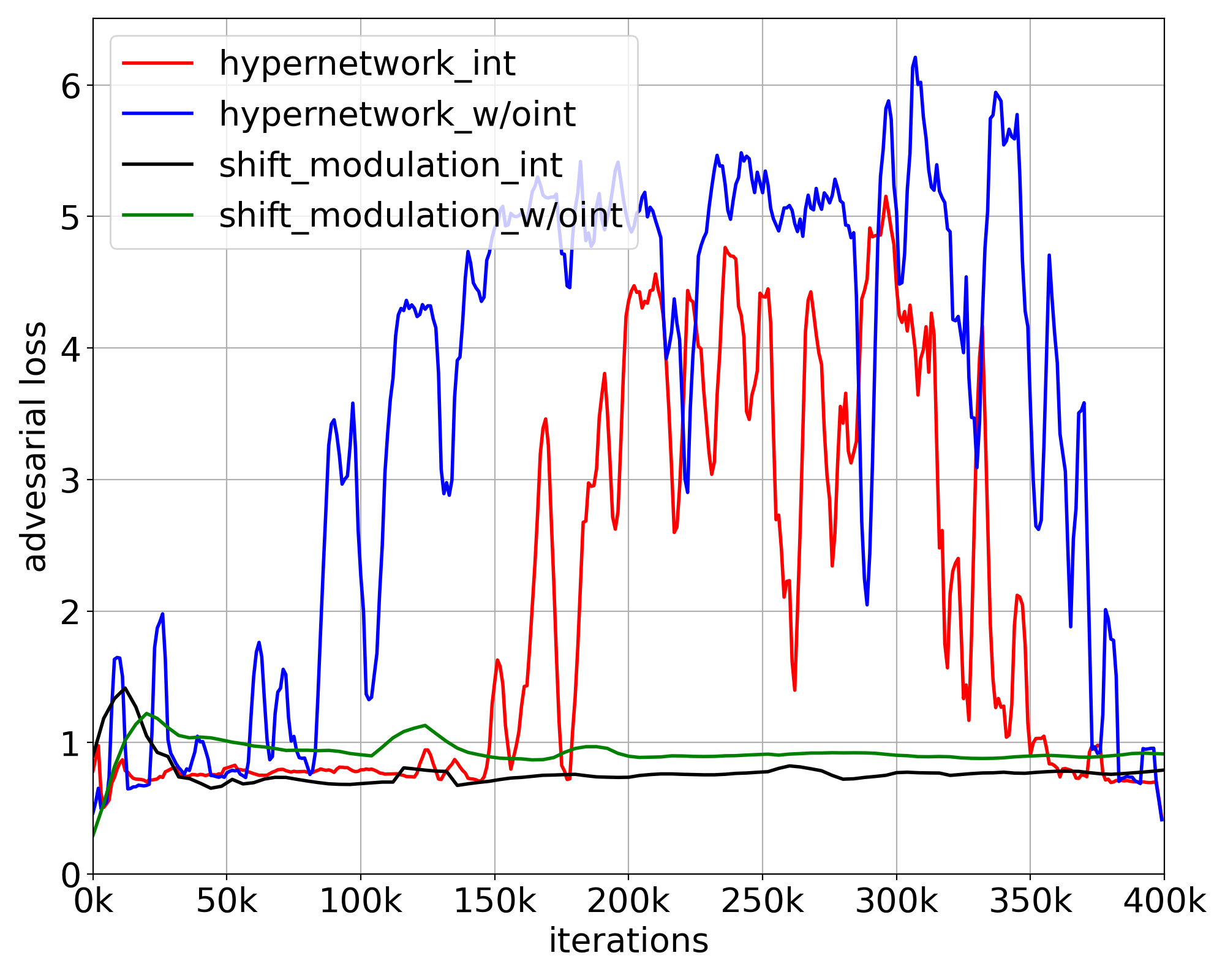} 
        \end{subfigure}
        \hspace{0.2em}
        \begin{subfigure}[h]{0.49\textwidth}
            \caption{Total generator loss curves}
            \includegraphics[width=\textwidth]{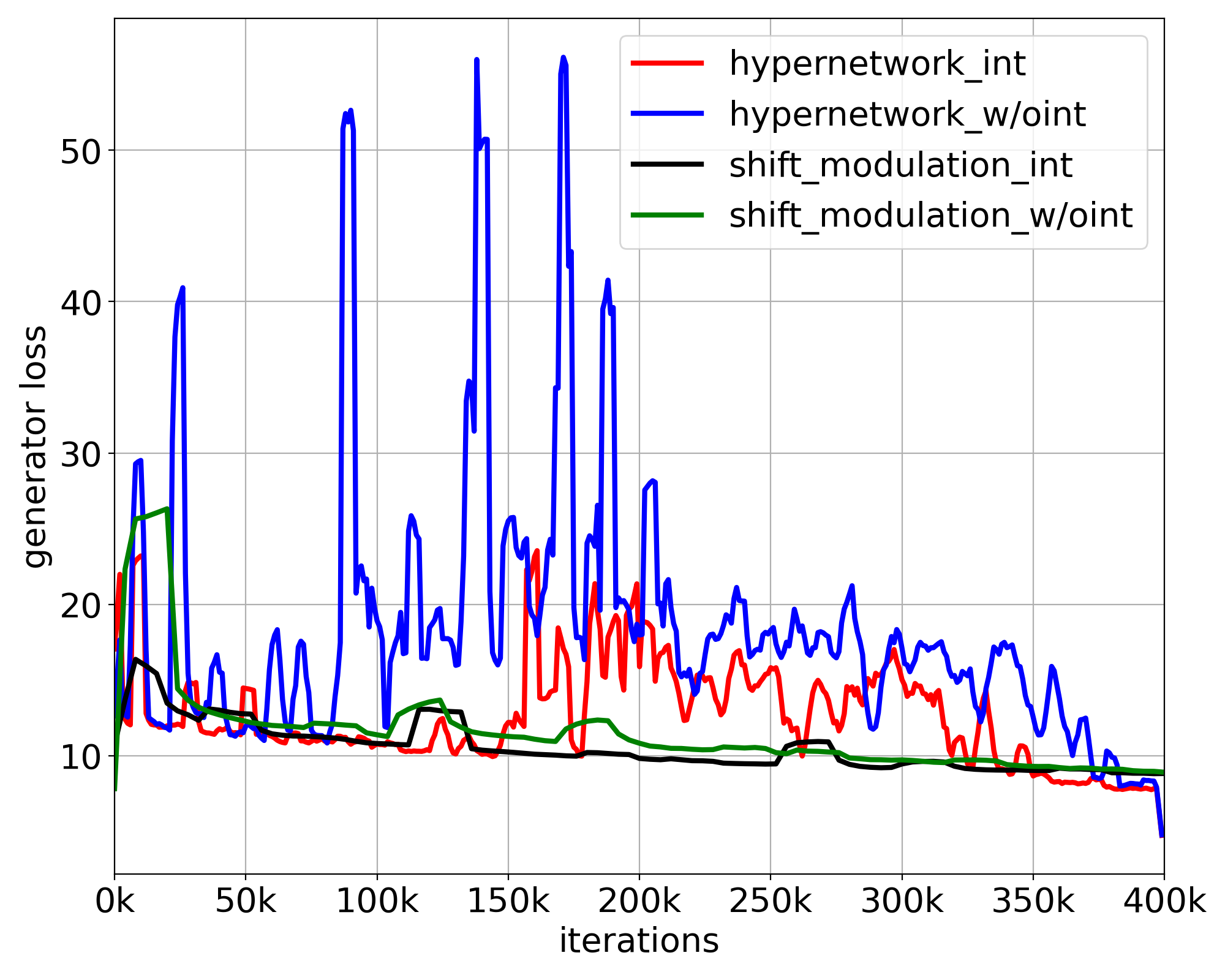}  
        \end{subfigure}
    \caption{Training loss curves of the ablated models. (a) adversarial loss (b) total generator loss including reconstruction loss, adversarial loss, feature matching loss, and latent code regularization. The models using shift modulation show more stable training loss against the models using a full hypernetwork.}
    \label{fig:loss_curve}
    \end{figure}

    We next demonstrated the stability of the models by comparing the loss curves of the ablated models. Both the adversarial loss $L_{\text{adv}}$ and the total loss $L$ are shown in Fig.~\ref{fig:loss_curve}. We observe that both losses of the models using the full hypernetwork fluctuated substantially and $L_{\text{adv}}$ increased midway through training. On the contrary, both loss curves of the models using shift modulation remained stable throughout the learning. The experiments suggested that by reducing the number of parameters generated, shift modulation is able to improve the stability of the image translation model.

\section{Discussion and conclusion}
\label{sec:conclusion}
In this work, we proposed CoNeS, a novel conditional neural fields-based model for MRI translation. We modeled the image translation problem using neural fields, which are conditioned on the source images, and learned latent codes through a coordinate-based network. The proposed model adapts the predicted neural fields by varying the latent codes across coordinates to ensure better local representations. We then introduced a shift modulation strategy for the conditioning to reduce the model complexity and stabilize the training. We compared the proposed model with state-of-the-art image translation models and our experiments showed that CoNeS performs better in the entire image scope as well as the tumor region, which is clinically relevant. Through visualization results, we also showed that the proposed method can reproduce more structural details, while the other methods' results are visually more blurry. The transformer-based model (ResViT) performed on par with the other methods in our experiments, where on natural images they have been reported to outperform those ~\citep{Kim2022Insta,shibasaki20224k}. Our datasets, however, are considerably smaller than what is used in the domain of natural images, while transformer-based models are considered data-demanding.

We performed a spectral analysis to demonstrate improvements in image translation when using neural fields. As expected, all the CNN-based models and ResViT, which is a hybrid transformer model containing transposed layers during decoding, were unable to reproduce high-frequency signals due to their spectral bias \citep{rahaman2019spectral,durall2020watch}. In contrast, the proposed model was able to preserve the high-frequency information and reconstruct the spectrum in the entire frequency range on both datasets. We also observed that ASAP-Net, a neural field-based benchmark, did not show consistent performance across the two datasets and could not reproduce the spectral distribution on the VS dataset either. These results are consistent with prior studies demonstrating that the full hypernetwork, in which all the parameters of the main network are generated, is sensitive to its initialization and difficult to optimize \citep{chang2019principled}. The ablation studies further indicated that compared to a full hypernetwork, the conditioning via shift modulation can make the training of neural fields more stable and maintain the representation capability. Furthermore, the results also showed that by introducing the adversarial loss, the predicted images are more realistic and contain more textural detail, although the quantitative metrics (SSIM and PSNR) are slightly lower than the model without the adversarial loss. The reason may be that SSIM and PSNR are not able to measure the benefits of the adversarial loss, which is in line with the conclusion in previous research \citep{liu2023one,dalmaz2022resvit}.

To evaluate the value of synthesized MRI in downstream analysis, we performed tumor segmentation experiments. We first demonstrated that dropping sequences during inference of a segmentation model can significantly damage the performance, which shows the complementary importance of multiple MRI sequences in segmentation. We next tested the segmentation model using different synthesized images and compared the results with the inference using incomplete input images. The experiments demonstrated that image translation models can significantly improve segmentation accuracy by replacing the missing input channel with synthesized images. Furthermore, the images generated by our proposed CoNeS model performed best among the state-of-the-art methods in most of the experiments, which is consistent with the visual improvement observed in the translation experiments. Nevertheless, we found that synthesized images cannot fully replace real images, and a baseline model trained on all real images performed best.

One limitation of our work is that in the clinic, the availability of MRI sequences may vary from patient to patient \citep{li2023brain}. The proposed model, however, cannot handle arbitrarily missing sequences, unless separate models are trained for each case. Further work would be adapting the proposed model to random incomplete MRI scans by incorporating techniques like learning disentangled representations \citep{shen2020multi} or latent representation fusion \citep{chartsias2017multimodal}. Moreover, the choice of the positional encoding frequency $m$ may bias the network to fit the signal of a certain bandwidth \citep{Wang2021SplinePE}. To ease the optimization and improve the generalization, it may be worthwhile to integrate periodic activation functions \citep{sitzmann2020implicit} in our design instead of positional encoding for better representation capability.  

In summary, we presented a neural fields-based model that synthesizes missing MRI from other sequences with excellent performance, which can be further integrated into the downstream analysis. All experiments showed improved performance compared to state-of-the-art translation models, while the spectrum analysis and ablation studies demonstrated the strengths of the proposed model over traditional CNN and neural fields models. Neural fields hold great promise in MRI translation to solve the missing MRI sequence problem in the clinic. 


\acks{This study was supported by the China Scholarship Council (grant 202008130140), and by an unrestricted grant of Stichting Hanarth Fonds, The Netherlands (project MLSCHWAN).}

%
\ethics{The work follows appropriate ethical standards in conducting research and writing the manuscript, following all applicable laws and regulations regarding the treatment of animals or human subjects.}

\coi{We declare we do not have conflicts of interest.}

\bibliography{references}



\end{document}